# Distributed and Optimal Reduced Primal-Dual Algorithm for Uplink OFDM Resource Allocation

Xiaoxin Zhang, Liang Chen, Jianwei Huang, Minghua Chen, and Yuping Zhao

***Abstract*—Orthogonal Frequency Division Multiplexing (OFDM) is the key component of many emerging broadband wireless access standards. The resource allocation in OFDM uplink, however, is challenging due to heterogeneity of users' Quality of Service requirements, channel conditions, and individual resource constraints. We formulate the resource allocation problem as a non-strictly convex optimization problem, which typically has multiple global optimal solutions. We then propose a reduced primal-dual algorithm, which is distributed, low in computational complexity, and probably globally convergent to a global optimal solution. The performance of the algorithm is studied through a realistic OFDM simulator. Compared with the previously proposed centralized optimal algorithm, our algorithm not only significantly reduces the message overhead but also requires less iterations to converge.**

## I. Introduction

Orthogonal Frequency Division Multiplexing (OFDM) is a promising technology for future broadband wireless networks. In OFDM, the entire frequency band is divided into a large number of subchannels, and network resource can be allocated flexibly over each of the subchannels. Because of this, OFDM enjoys several key advantages such as robustness against intersymbol interference and multipath fading, and no need for complex equalizations.

In this paper, we consider the resource allocation in a single cell OFDM uplink system, where multiple end users transmit data to the same base station. This is motivated by several practical wireless systems, such as WiMAX/802.16e, LTE for 3GPP, and UMB for 3GPP2. Given the channel conditions of the users at a particular time, we need to determine which subset of users to schedule (i.e., allowed to transmit with positive rates), how to match the subchannels with the scheduled users, and the power allocation across these subchannels.

We formulate the resource allocation problem as a weighted rate maximization problem, which is motivated by the gradient-based scheduling framework in [1]–[3]. This problem, however, is quite challenging to solve due to the heterogeneity of users' Quality of Service requirements, channel conditions, and individual resource constraints. In this paper, we propose the first distributed primal-dual algorithm in literature that achieves the optimal resource allocation in uplink OFDM systems. Our key contributions are:

- *Realistic OFDM model*: we consider several important practical constraints such as self-noise and per subchannel per user SNR limits. These constraints are typically ignored in previous work but have significant impacts on the optimal solution as well as algorithm design.
- *Optimal algorithm with global convergence*: the proposed algorithm is provably globally convergent to one of the global optimal solutions of the resource allocation problem, despite of non-strict convexity of the problem under which setting primal-dual algorithms may not be able to converge (e.g., [4]–[6]).
- *Distributed algorithm with scalable performance*: the proposed algorithm is distributed, requires simple local updates, demands only limited message passing, and is highly scalable with network size.
- *Simple algorithm with fast convergence*: the proposed algorithm only updates a subset of all decision variables. Compared with the previously proposed centralized algorithm, our algorithm reduces complexity and requires less iterations to converge.

In Section II we present the network model and problem statement. The reduced primal-dual algorithm is proposed in Section III, and its convergence behavior is analyzed in Section IV. Simulations results on convergence and optimality are given in Section V, and we conclude in Section VI.

### A. Related Work

Most previous work on resource allocation in OFDM systems focused on the downlink case, where the base station sends traffic to multiple end users with a total power constraint. Compared with the uplink case, the optimization problem in the downlink case is easier to solve and a centralized algorithm is reasonable to implement (e.g., [7]). Due to different resource constraints, however, the algorithms proposed for the downlink case can not be directly applied to the uplink case.

Uplink OFDM resource allocation only receives attention recently, e.g., [8]–[14]. In [8], the problem was formulated in the framework of Nash Bargaining with a focus of fair resource allocation. The authors of [9] proposed a heuristic algorithm that tries to minimize each user's transmission power while satisfying the individual rate constraints. In [10], the author considered the sum-rate maximization problem and derived algorithms based on Rayleigh fading on each subchannel. The authors in [11]–[14] proposed several heuristic algorithms to solve a problem similar as the one considered here with

X. Zhang and Y. Zhao are with the State Key Laboratory of Advanced Optical Communication Systems & Networks, School of Electronics Engineering and Computer Science, Peking University, Beijing 100871, P.R.China, e-mail: christine.xiaoxin@gmail.com, yuping.zhao@pku.edu.cn. L. Chen, J. Huang and M. Chen are with the Department of Information Engineering, The Chinese University of Hong Kong, Hong Kong, e-mail: {lchen6,jwhuang,minghua}@ie.cuhk.edu.hk.

additional integer channel allocation constraints. None of the previous literature focused on solving the uplink resource allocation problem optimally.

In a recent work [15], the authors proposed a dual-based centralized algorithm to solve the resource allocation problem. In the centralized algorithm, the base station needs to collect the complete network information (e.g., users' weights, resource constraints, and uplink channel conditions) before the algorithm starts. It also needs to have sophisticated computational capability to perform the dual-based algorithm. Finally, recovering the primal optimal solution after the dual algorithm converges is not trivial. Our algorithm overcomes all three bottlenecks since it is distributed, simple, and provably converges to the optimal primal variables.

This paper lies in the area of using primal-dual algorithms for non-strictly convex optimization problems. Under the non-strict convex setting, primal-dual algorithms in general may not converge to the optimal solution, as observed in literature [4], [6]. In [6], the authors characterized the convergence behavior of a certain primal-dual algorithm in a P2P setting, and gave a sufficient condition under which the algorithm is guaranteed to converge. Our analysis is motivated by the convergence proof in [6], but we deal with a different set of challenges in a totally different problem setting.

This work is a generalization of our previous work [5], where we presented some preliminary results on the primal-dual algorithm for a simple OFDM model. Here we consider a much complicated model including various constraints such as self-noise and SNR limitations. Moveover, we design a reduced primal-dual algorithm by considering the relationship that the variables should satisfy at the optimal solution. In other words, the dynamics of our newly proposed algorithm is constrained by a high dimension nonlinear manifold. All the above require different techniques in terms of proving the optimality and convergence. We also perform extensive simulations to illustrate the performance of the proposed algorithm together with comparison with the previous dual-based centralized algorithm.

## II. Problem Statement

We consider a single OFDM cell, where there is a set $\mathcal{M} = \{1, \dots, M\}$ of users transmitting to the same base station. Each user $i \in \mathcal{M}$ has a priority weight $w_i$.[1] The total frequency band is divided into a set $\mathcal{N} = \{1, \dots, N\}$ of subchannels (e.g., tones/carriers). A user $i \in \mathcal{M}$ can transmit over a subset of the subchannels, with transmit power $p_{ij}$ over subchannel $j \in \mathcal{N}$ and the total power upper bound, i.e., $\sum_j p_{ij} \leq P_i$. For channel $j$, it is allocated to user $i$ with fraction $x_{ij}$, and the total allocation across all users should be no larger than 1, i.e., $\sum_i x_{ij} \leq 1$.

We define $e_{ij}$ as the received signal-to-noise ratio (SNR) per unit power for user $i$ on subchannel $j$. As in [7], this model can also incorporate various sub-subchannelization schemes where the resource allocation is performed in terms of sets of frequency bands in the frequency domain or with a granularity of multiple symbols in the time domain. We further assume that the channel conditions do not change within the time of interests, i.e., we are looking at a resource allocation period smaller than the channel coherence time[2].

TABLE I

Key Notations

| Notation | Physical Meaning |
|---|---|
| $N$ | total number of subchannels |
| $\mathcal{N}$ | set of all subchannels |
| $M$ | total number of users |
| $\mathcal{M}$ | set of all users |
| $i$ | user index |
| $j$ | subchannel index |
| $w_i$ | user $i$'s (dynamic) weight |
| $e_{ij}$ | normalized SNR on subchannel $j$ for user $i$ |
| $p_{ij}$ | power allocated on subchannel $j$ for user $i$ |
| $x_{ij}$ | fraction of subchannel $j$ allocated to user $i$ |
| $P_i$ | maximum transmit power for user $i$ |
| $s_{ij}$ | maximum SNR coefficient on subchannel $j$ for user $i$ |
| $\beta$ | self-noise coefficient |

The key notations used throughout this paper are listed in Table I. We use bold symbols to denote vectors and matrices of these quantities, e.g., $\boldsymbol{w} = \{w_i, \forall i\}$, $\boldsymbol{e} = \{e_{ij}, \forall i, j\}$, $\boldsymbol{p} = \{p_{ij}, \forall i, j\}$, and $\boldsymbol{x} = \{x_{ij}, \forall i, j\}$.

Our objective is to maximize the weighted sum of the users' rates over the feasible rate-region defined as follows,

$$\mathcal{R}(\boldsymbol{e}) = \left\{ \boldsymbol{r} \in \Re_+^M : r_i = \sum_{j \in \mathcal{N}} x_{ij} \log\left(1 + \frac{p_{ij} e_{ij}}{x_{ij} + \beta p_{ij} e_{ij}},\right),\ \forall i \in \mathcal{M} \right\}, \tag{1}$$

where $r_i$ is the achievable rate of user $i$[3] and $(\boldsymbol{x}, \boldsymbol{p}) \in \mathcal{X}$ are chosen subject to

$$\sum_i x_{ij} \leq 1, \forall j \in \mathcal{N}, \tag{2}$$

$$\sum_j p_{ij} \leq P_i, \forall i \in \mathcal{M}, \tag{3}$$

and the set

$$\mathcal{X} := \left\{ (\boldsymbol{x}, \boldsymbol{p}) \geq \mathbf{0} : 0 \leq x_{ij} \leq 1, 0 \leq p_{ij} \leq \frac{x_{ij} s_{ij}}{e_{ij}}, \forall i, j \right\}. \tag{4}$$

Here, $\beta$ represents the level of "self-noise" because of imperfect carrier synchronization or inaccurate channel estimation (e.g. [16], [17]). $s_{ij}$ is a maximum SNR constraint on subchannel $j$ for user $i$, modeling for example limited choices of available modulation and coding schemes[4].

In many OFDM standards, $x_{ij}$ is constrained to be an integer, in which case we add the additional constraint $x_{ij} \in \{0, 1\}$ for all $i, j$. The integer constraint makes the resource allocation very difficult to solve, and various heuristic algorithms dealing with such constraint are proposed in [12]–[15]. In this paper, we will ignore this integer constraint and focus on the rate region defined by (1) to (4). The corresponding solution typically contains fractional values of $x_{ij}$'s. There are several practical methods of achieving these fraction allocations. For example, if resource allocation is done in blocks of OFDM symbols, then fractional values of $x_{ij}$ can be implemented by time-sharing the symbols in a block. Likewise, if the number of subchannels are large enough so that the channel conditions do not change dramatically among adjacent subchannels, then the fractional values of $x_{ij}$'s can also be implemented by frequency sharing (e.g., [18]).

[1] We will later discuss how these weights are derived in practice.

[2] This is particularly suitable for fixed broadband wireless access (part of the IEEE 802.16 standard), where users are relatively static.

[3] For notation simplicity we normalize the bandwidth to be 1 during the analysis. A realistic value will be considered in the simulations (Section V).

[4] We restrict our attention to the interesting "resource constrained" case, where constraints (3) are tight for all users at an optimal solution with the maximum SNR constraints taken into consideration.

In this paper, we focus on solving the following problem

$$\max_{\boldsymbol{r}\in\mathcal{R}(\boldsymbol{e})} \sum_i w_i r_i, \tag{5}$$

where rate $r_i$ and rate region $\mathcal{R}(\boldsymbol{e})$ are given by (1).

Finally, we note that the priority weights $w_i$'s are motivated by the gradient-based scheduling framework presented in [1]–[3]. Let's assume each user $i$ has a utility function $U_i(W_{i,t})$ depending on its average throughput $W_{i,t}$ up to time $t$. It has been shown that in order to maximize the total network utility $\sum_i U_i(W_{i,t})$ in the long run, it is enough to solve Problem (5) during each time slot $t$ with $w_i = \partial U_i(W_{i,t})/\partial W_{i,t}$.

## III. A Reduced Primal-Dual Algorithm

We can rewrite problem (5) in variables $\boldsymbol{x}$ and $\boldsymbol{p}$ as follows:

*Problem 1 (Weighted Rate Maximization):*

$$\max_{(\boldsymbol{x},\boldsymbol{p})\in\mathcal{X}} \sum_{i\in\mathcal{M}} w_i \sum_{j\in\mathcal{N}} x_{ij} \log\left(1 + \frac{p_{ij}e_{ij}}{x_{ij} + \beta p_{ij}e_{ij}}\right), \tag{6}$$

subject to the per subchannel assignment constraints in (2) and the per user power constraints in (3). Set $\mathcal{X}$ is given in (4).

Although the objective function in (6) is concave, its derivative is not well defined at the origin ($\boldsymbol{x} = 0, \boldsymbol{p} = 0$). This motivates us to look at the following $\epsilon$-relaxed version of Problem 1:

*Problem 2 ($\epsilon$-relaxed Weighted Rate Maximization):*

$$\max_{(\boldsymbol{x},\boldsymbol{p})\in\mathcal{X}} \sum_{i\in\mathcal{M}} w_i \sum_{j\in\mathcal{N}} (x_{ij} + \epsilon_{ij}) \log\left(1 + \frac{p_{ij}e_{ij}}{x_{ij} + \beta p_{ij}e_{ij} + \epsilon_{ij}}\right), \tag{7}$$

where constants $\epsilon_{ij}$ take small positive value for all $i$ and $j$. The constraint set remains the same as in Problem 1.

By such relaxation, the objective function in (7) now has derivative defined everywhere in the constraint set $\mathcal{X}$. Thanks to the continuity of the objective function, the optimal value to Problem 2 can be arbitrarily close to that of Problem 1, if $\boldsymbol{\epsilon} = [\epsilon_{ij}, \forall i, j]$ is chosen to be small enough.

The constraint set of Problem 2 is convex, and the objective function in (7) is continuous and *non-strictly* concave. As such, Problem 2 has multiple optimal solutions, and there is no duality gap between it and its dual problem.

The existence of derivatives allows us to write down a primal-dual algorithm to pursue the optimal solution to Problem 2. The Lagrangian for Problem 2 is as follows,

$$L(\boldsymbol{\lambda}, \boldsymbol{\mu}, \boldsymbol{x}, \boldsymbol{p}) := \sum_{i,j} w_i(x_{ij} + \epsilon_{ij}) \log(1 + \frac{p_{ij}e_{ij}}{x_{ij} + \beta p_{ij}e_{ij} + \epsilon_{ij}}) + \sum_i \lambda_i(P_i - \sum_j p_{ij}) + \sum_j \mu_j(1 - \sum_i x_{ij}). \tag{8}$$

By strong duality theorem, the optimal primal and dual solutions must satisfy KKT conditions, i.e., for all $i$ and $j$,

$$\mu_j \geq 0, \ \sum_i x_{ij} \leq 1, \ \mu_j(\sum_i x_{ij} - 1) = 0, \tag{9}$$

$$\lambda_i \geq 0, \ \sum_j p_{ij} \leq P_i, \ \lambda_i(\sum_j p_{ij} - P_i) = 0, \tag{10}$$

$$x_{ij} \geq 0, \ 0 \leq p_{ij} \leq \frac{x_{ij}s_{ij}}{e_{ij}}, \tag{11}$$

$$x_{ij}(f_{ij}(x_{ij}, p_{ij}) - \mu_j) \leq 0, \tag{12}$$

$$p_{ij}(g_{ij}(x_{ij}, p_{ij}) - \lambda_i) \leq 0, \tag{13}$$

where $f_{ij}(\cdot)$ and $g_{ij}(\cdot)$ are gradients of the objective function in (7) with respect to $x_{ij}$ and $p_{ij}$, respectively, and are given by

$$f_{ij}(x_{ij}, p_{ij}) = w_i \log(1 + \frac{p_{ij}e_{ij}}{x_{ij} + \beta p_{ij}e_{ij} + \epsilon_{ij}}) - \frac{w_i(x_{ij} + \epsilon_{ij})p_{ij}e_{ij}}{(x_{ij} + \beta p_{ij}e_{ij} + \epsilon_{ij})[x_{ij} + (\beta + 1)p_{ij}e_{ij} + \epsilon_{ij}]},$$

and

$$g_{ij}(x_{ij}, p_{ij}) = \frac{w_i e_{ij}(x_{ij} + \epsilon_{ij})^2}{(x_{ij} + \beta p_{ij}e_{ij} + \epsilon_{ij})[x_{ij} + (\beta + 1)p_{ij}e_{ij} + \epsilon_{ij}]}.$$

The last two KKT conditions in (12) and (13) become equalities if $x_{ij} > 0$ and $p_{ij} > 0$, respectively. It can be verified that the optimal solutions of Problem 2, satisfying above KKT conditions, are exactly the saddle points of the Lagrangian function in (8). Since the primal problem has at least one solution, the saddle point exists.

For notation simplicity, we define $(a)^+ = \max(a, 0)$,

$$(a)^+_b = \begin{cases} a, & b > 0, \\ \max(a, 0), & \text{otherwise}, \end{cases}$$

and

$$(a)^{\pm}_{b,c} = \begin{cases} a, & 0 < b < c, \\ \max(a, 0), & b \leq 0, \\ \min(a, 0), & \text{otherwise}. \end{cases}$$

To pursue saddle points of the Lagrangian function, we start by considering a *standard-form* primal-dual algorithm, then we derive a new *reduced* primal-dual algorithm which has less complexity and faster convergence speed than the standard one.

Motivated by the work in [5], we consider the following standard-form primal-dual algorithm: $\forall i, j$,

$$\dot{x}_{ij} = k^x_{ij}(f_{ij}(x_{ij}, p_{ij}) - \mu_j)^+_{x_{ij}}, \tag{14}$$
$$\dot{p}_{ij} = k^p_{ij}(g_{ij}(x_{ij}, p_{ij}) - \lambda_i)^{\pm}_{p_{ij}, \frac{x_{ij}s_{ij}}{e_{ij}}}, \tag{15}$$
$$\dot{\mu}_j = k^{\mu}_j(\sum_i x_{ij} - 1)^+_{\mu_j}, \tag{16}$$
$$\dot{\lambda}_i = k^{\lambda}_i(\sum_j p_{ij} - P_i)^+_{\lambda_i}, \tag{17}$$

where $k^x_{ij}, k^p_{ij}, k^{\mu}_j$ and $k^{\lambda}_i$ are constants representing adaption rates. Here the derivatives are defined with respect to time.

We call a point $(\boldsymbol{x}, \boldsymbol{p}, \boldsymbol{\mu}, \boldsymbol{\lambda})$ an *equilibrium* of the standard-form algorithm if and only if the corresponding derivatives in (14) to (17) are zero for all $i$ and $j$. We can show that the set of equilibria of the standard-form primal-dual algorithm is equivalent to the set of global optimal solutions of Problem 2.

In the standard-form primal-dual algorithm, all variables $x_{ij}$, $p_{ij}$, $\mu_j$, and $\lambda_i$ are dynamically adapted. This might lead to performance concerns in terms of high complexity and slow convergence. One way to address this concern is to reduce the number of dynamically adapting variables. We achieve this goal by constraining the algorithm trajectories onto a manifold which includes all optimal primal and dual solutions.

We study the following manifold by setting (15) to zero, i.e. $\forall i, j$,

$$0 = (g_{ij}(x_{ij}, p_{ij}) - \lambda_i)^{\pm}_{p_{ij}, \frac{x_{ij}s_{ij}}{e_{ij}}}, \tag{18}$$

which in turn implies

$$\begin{cases} g_{ij}(x_{ij}, p_{ij}) = \lambda_i, & \text{if } 0 < p_{ij} \le \frac{x_{ij}s_{ij}}{e_{ij}}, \\ g_{ij}(x_{ij}, p_{ij}) \le \lambda_i, & \text{if } p_{ij} = 0. \end{cases} \tag{19}$$

Clearly, the optimal primal and dual solutions must lie on the above manifold.

After simplification we get the following expression of the manifold:

$$p_{ij} = \min\left\{\frac{x_{ij}s_{ij}}{e_{ij}},\ h_{ij}\right\}, \tag{20}$$

where $h_{ij}$ is denoted by

$$h_{ij} = \left(\frac{\sqrt{1 + 4\beta(\beta+1)\frac{w_i e_{ij}}{\lambda_i}} - (2\beta+1)}{2\beta(\beta+1)e_{ij}}(x_{ij} + \epsilon_{ij})\right)^+_{w_i e_{ij} - \lambda_i}$$

when $\beta \neq 0$, and

$$h_{ij} = \left(\frac{w_i e_{ij} - \lambda_i}{\lambda_i e_{ij}}(x_{ij} + \epsilon_{ij})\right)^+_{w_i e_{ij} - \lambda_i}$$

when $\beta = 0$.

Substituting (20) into (14) to (17), we obtain a new reduced primal-dual algorithm as follows:

***Algorithm RPD: Reduced Primal-Dual Algorithm***

$$\dot{x}_{ij} = k^x_{ij}(f_{ij}(x_{ij}, p_{ij}) - \mu_j)^+_{x_{ij}}, \tag{21}$$
$$\dot{\mu}_j = k^{\mu}_j(\sum_i x_{ij} - 1)^+_{\mu_j}, \tag{22}$$
$$\dot{\lambda}_i = k^{\lambda}_i(\sum_j p_{ij} - P_i)^+_{\lambda_i}, \tag{23}$$
$$p_{ij} = \min\left\{\frac{x_{ij}s_{ij}}{e_{ij}},\ h_{ij}\right\}. \tag{24}$$

*Proposition 1:* The set of equilibria of Algorithm RPD is the same as the set of global optimal solutions of Problem 2.

This means that if the reduced primal-dual algorithm converges, it reaches a global optimal solution of the $\epsilon$-relaxed weighted rate maximization problem.

Compared to the standard-form algorithm in (14) to (17) in which $\boldsymbol{p}$ is dynamically adapted, $\boldsymbol{p}$ in the new Algorithm RPD is directly computed from $\boldsymbol{x}$ and $\boldsymbol{\lambda}$. Consequently, the reduced Algorithm RPD has less dynamically adapting variables than the standard-form algorithm, and hence is less complex and is expected to converge faster.

Similar as the standard-form algorithm, the reduced Algorithm RPD can also be implemented in a distributed fashion by end users and the base station. A end user $i$ is responsible of updating $x_{ij}$'s as well as dual variable $\lambda_i$ locally. During each iteration, it sends the latest values of $x_{ij}$'s to the base station, but not the $p_{ij}$'s or $\lambda_i$. The base station is responsible for updating dual variables $\mu_j$'s for all subchannels and broadcasting to the users. In particular, the base station does not need to know users' channel conditions, power constraints, or priority weights. Both the communication complexity and computation complexity per iteration are $O(MN)$.

Despite of its advantage, we need to first show that trajectories of Algorithm RPD converge to its equilibria before we can apply it to solve Problem 2.

## IV. Convergence of the Reduced Primal-Dual Algorithm

In Algorithm RPD, we remove power allocation variables $p_{ij}$'s from the dynamically adapting variables. Now, the difficulty is how to ensure the global convergence of Algorithm RPD. The key challenge of the convergence proof is the *non-strict concavity* of the objective function in (7). It has been well observed in the literature that although primal-dual algorithms can globally converge to the optimal solution of a strictly concave optimization problem, they may oscillate indefinitely and fail to converge when applying to a non-strictly concave optimization problem, even setting small enough update step-sizes [4]–[6], [19].

In this section, we successfully prove the sufficient conditions under which Algorithm RPD can globally and asymptotically converge to a global optimal solution of Problem 2. The proof is organized as follows. We first show in Theorem 1 that the trajectories of Algorithm RPD converge to an invariant set containing all global optimal solutions of Problem 2. Then in Lemma 1 we show that the trajectories of Algorithm RPD might form limit cycles within the invariant set if this set contains non-optimal solutions. In Theorem 2 we present the conditions under which the invariant set only contains global optimal solutions of Problem 2 without limit cycles. Finally we show in Corollary 1 how to choose the update stepsizes to satisfy the sufficient conditions in Theorem 2.

*Theorem 1:* All trajectories of Algorithm RPD converge to an invariant set $V_0$ globally and asymptotically. Furthermore, let $(\boldsymbol{x}^*, \boldsymbol{p}^*, \boldsymbol{\mu}^*, \boldsymbol{\lambda}^*)$ be a global optimal solution of Problem 2 and $(\boldsymbol{x}, \boldsymbol{p}, \boldsymbol{\mu}, \boldsymbol{\lambda})$ be any point in set $V_0$, the following is true for all $i$ and $j$,

1) $(\boldsymbol{x}^*, \boldsymbol{p}^*, \boldsymbol{\mu}^*, \boldsymbol{\lambda}^*)$ is contained in $V_0$;
2) $\mu_j$ is nonzero only if $\sum_i x^*_{ij} = 1$;
3) $\sum_j p^*_{ij} = P_i$, and $\lambda_i$ is a positive constant;
4) over set $V_0$, $f_{ij}(x_{ij}, p_{ij}) = f_{ij}(x^*_{ij}, p^*_{ij}) = \mu^*_j$, and $g_{ij}(x_{ij}, p_{ij}) = g_{ij}(x^*_{ij}, p^*_{ij}) = \lambda^*_i$;
5) $\frac{p_{ij}}{x_{ij}+\epsilon_{ij}} = \frac{p^*_{ij}}{x^*_{ij}+\epsilon_{ij}}$.

The proof of Theorem 1 can be found in Appendix A. Results 1) to 4) will be used in later analysis.

Result 5) of Theorem 1 is of independent interest. It implies that although there can be multiple global optimal solutions to Problem 2, the effective SNR achieved by user $i$ on subchannel $j$ is the same in all solutions.

Although all trajectories of Algorithm RPD may converge to the desired equilibria in $V_0$, they may also converge to non-equilibrium points in $V_0$ (if there are any). We now study the conditions for $V_0$ to contain *only* the desired equilibria, under which Theorem 1 guarantees the convergence of Algorithm RPD to a global optimal solution of Problem 2.

Plugging result 4) of Theorem 1 into Algorithm RPD, and recalling that $M$ is the total number of users and $N$ is the total number of subchannels, we find that $V_0$ is exactly the set that contains all trajectories of the following linear system in (25) to (27) over set $\{\boldsymbol{x} \geq 0, \boldsymbol{\mu} \geq 0\}$.

$$\dot{\boldsymbol{x}} = K^x A_1^T \boldsymbol{\mu}^* - K^x A_1^T \boldsymbol{\mu}, \quad (25)$$

$$\dot{\boldsymbol{\mu}} = K^\mu A_1 \boldsymbol{x} - K^\mu \mathbf{1}, \quad (26)$$

$$\dot{\boldsymbol{\lambda}} = K^\lambda A_2 B(\boldsymbol{x} + \boldsymbol{\epsilon}) - K^\lambda \boldsymbol{P} = 0, \quad (27)$$

where $K^x$ is an $MN \times MN$ diagonal matrix with diagonal terms equal to $k^x_{ij}$'s, $K^\mu$ is an $N \times N$ diagonal matrix with diagonal terms equal to $k^\mu_j$'s, and $K^\lambda$ is an $M \times M$ diagonal matrix with diagonal terms equal to $k^\lambda_i$'s. $B$ is an $MN \times MN$ diagonal matrix given by $B = diag(b_{ij},\ \forall i, j)$, where $b_{ij} = \frac{p^*_{ij}}{x^*_{ij}+\epsilon_{ij}}$. The matrix $A_1 = [I_N, \cdots, I_N]$ and has a dimension of $N \times MN$, where $I_N$ is an identity matrix with dimension $N$. The matrix $A_2$ has a dimension of $M \times MN$ and is given by

$$A_2 = \begin{bmatrix} \mathbf{1}_{1\times N} & 0 & \cdots & 0 \\ 0 & \mathbf{1}_{1\times N} & \cdots & 0 \\ \vdots & \vdots & \ddots & \vdots \\ 0 & 0 & \cdots & \mathbf{1}_{1\times N} \end{bmatrix},$$

where $\mathbf{1}_{1\times N}$ is an all one vector with dimension 1 by $N$.

For the above linear system, we have the following observations.

*Lemma 1:* For the linear system in (25) to (27), we have

1) every order Lie derivative of $A_2 Bx$ is constant, that is $\forall n \geq 0$:
$$\frac{d^n}{dt^n} A_2 B\boldsymbol{x} = \text{constant},$$
where $t$ denotes time;
2) starting from a non-equilibrium point, trajectories of $\boldsymbol{x}$ and $\boldsymbol{\mu}$, following (25) and (26) respectively, do not converge and form limit cycles.

The proof of Lemma 1 can be found in Appendix B.

Result 2 in Lemma 1 indicates that if $(x, \mu)$ starts from a non-equilibrium point, they will keep oscillating and never converge. Therefore, one way to guarantee the system convergence is to make sure that the invariant set $V_0$ only contains equilibria points (i.e., global optimal solutions). Result 1 in Lemma 1 states that every order Lie derivative of $A_2 B\boldsymbol{x}$ is constant. By linear system theory, if the system state $\boldsymbol{\mu}$ is completely observable from $A_2 B\boldsymbol{x}$, then constant $A_2 B\boldsymbol{x}$ will lead to $\dot{\boldsymbol{\mu}}$ equal to 0 and $\boldsymbol{\mu}$ being constant. When $\boldsymbol{\mu}$ is constant, $\dot{\boldsymbol{x}}$ is constant according to (25). Combining with the constraint that $\boldsymbol{x} \geq 0$ and $A_2 B\boldsymbol{x}$ is constant, we can show that $\dot{\boldsymbol{x}}$ is also zero if $\dot{\boldsymbol{\mu}}$ is zero over the set $\{\boldsymbol{x} \geq 0, \boldsymbol{\mu} \geq 0\}$.

In the following theorem, we state conditions for $\boldsymbol{\mu}$ to be completely observable from $A_2 B\boldsymbol{x}$, and summarize its consequence on convergence of Algorithm RPD.

*Theorem 2:* All trajectories of Algorithm RPD converge globally and asymptotically to the system equilibria if the following condition holds:

$$\begin{bmatrix} A_2 B K^x A_1^T \\ K^\mu A_1 K^x A_1^T - \sigma I \end{bmatrix} \text{ has rank } N, \quad (28)$$

where $\sigma$ denotes any eigenvalue of matrix $K^\mu A_1 K^x A_1^T$.

The proof of Theorem 2 can be found in Appendix C.

For the problem we studied in this paper, we can choose properly the update stepsizes of the algorithm to satisfy the condition in (28).

*Corollary 1:* The condition (28) in Theorem 2 is satisfied if both of the following are true

- $K^x = kI$ (diagonal terms of $K^x$ take the same value $k$);
- all diagonal elements of $K^\mu$ take different values.

The proof of Corollary 1 can be found in Appendix D.

In this section, we have investigated the convergence of Algorithm RPD by combining both La Salle principle from nonlinear stability theory and complete observability from linear system theory. The proof shows that Algorithm RPD can globally and asymptotically converge to one of the global optimal solutions of Problem 2 when satisfying the conditions in Corollary 1. The convergence proof of Algorithm SPD is similar and will not be presented here due to space limitation.

## V. Simulation Results

We show the convergence and optimality of the reduced primal-dual algorithm in Algorithm RPD over a realistic OFDM uplink simulator. Each user's subchannel gains $e_{ij}$'s are the product of two terms: a constant location-based term picked using an empirically obtained distribution, and a fast fading term generated using a block-fading model and a standard mobile delay-spread model with a delay spread of 10$\mu$sec. The system bandwidth is 5MHz consisting of 512 OFDM tones, which is further grouped into 64 subchannels.[5] The symbol duration is 100$\mu$sec with a cyclic prefix of 10$\mu$sec.

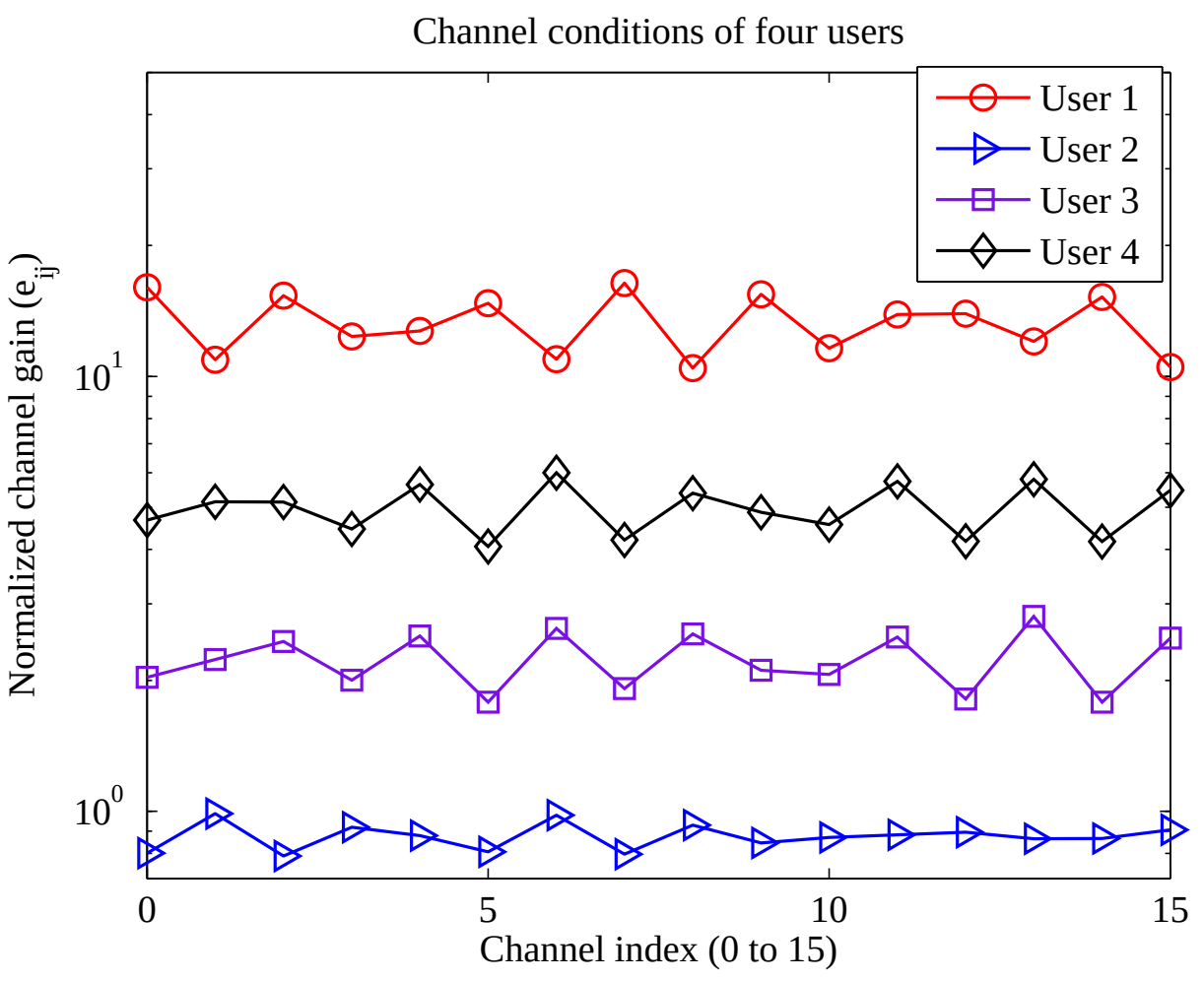

Fig. 1. Channel conditions of 4 users and 16 channels

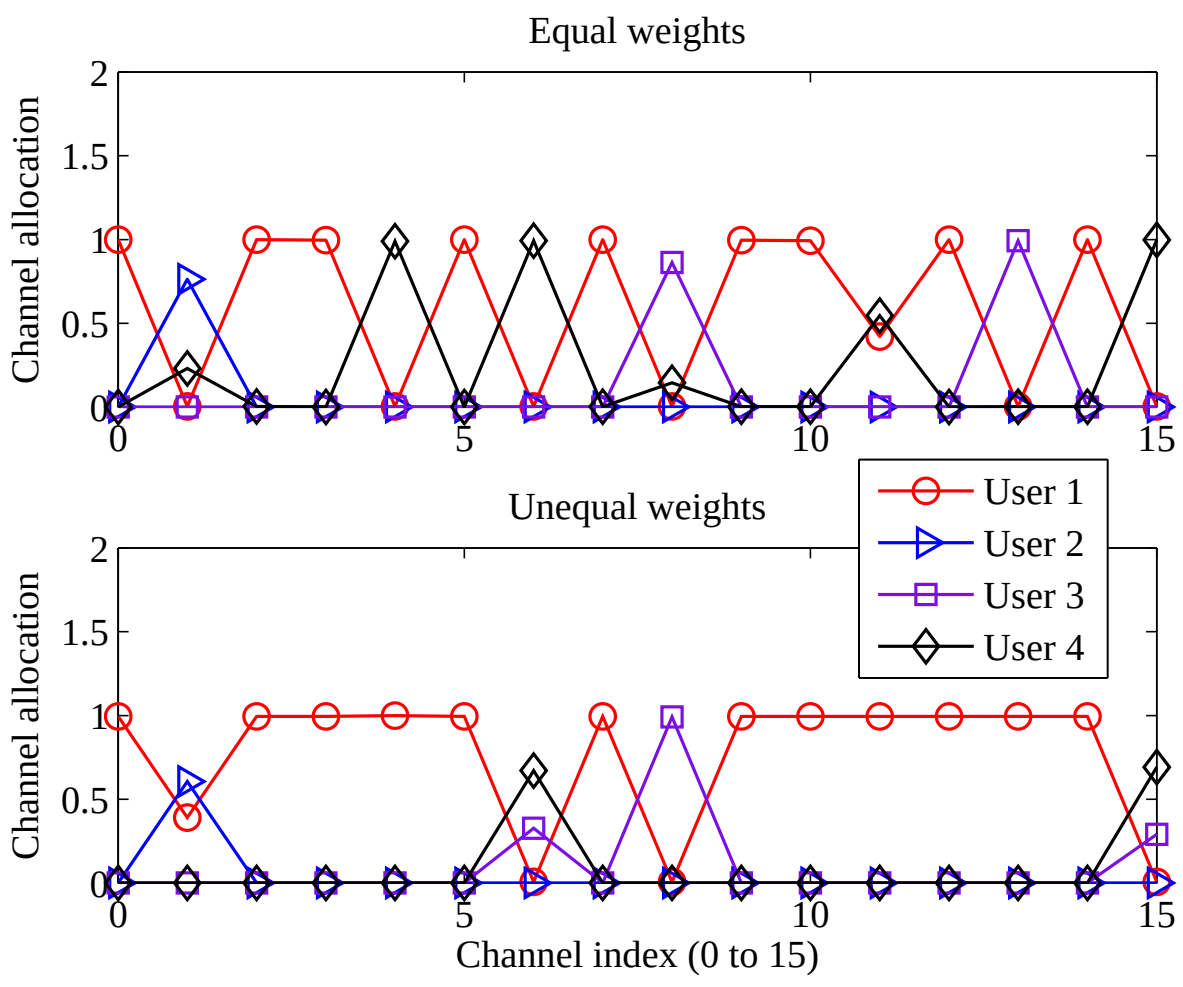

Fig. 2. Optimal channel allocation under Algorithm RPD

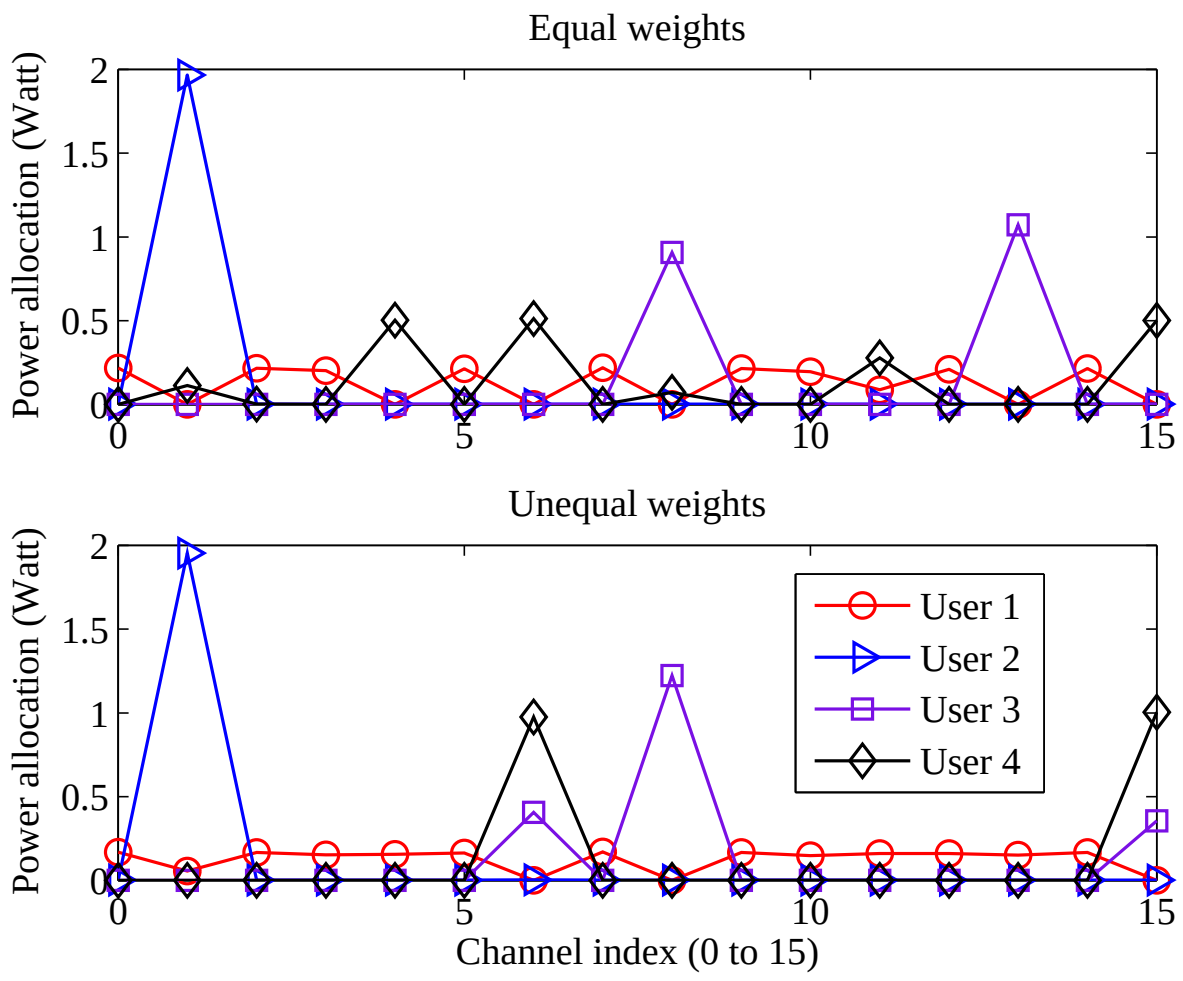

Fig. 3. Optimal power allocation under Algorithm RPD

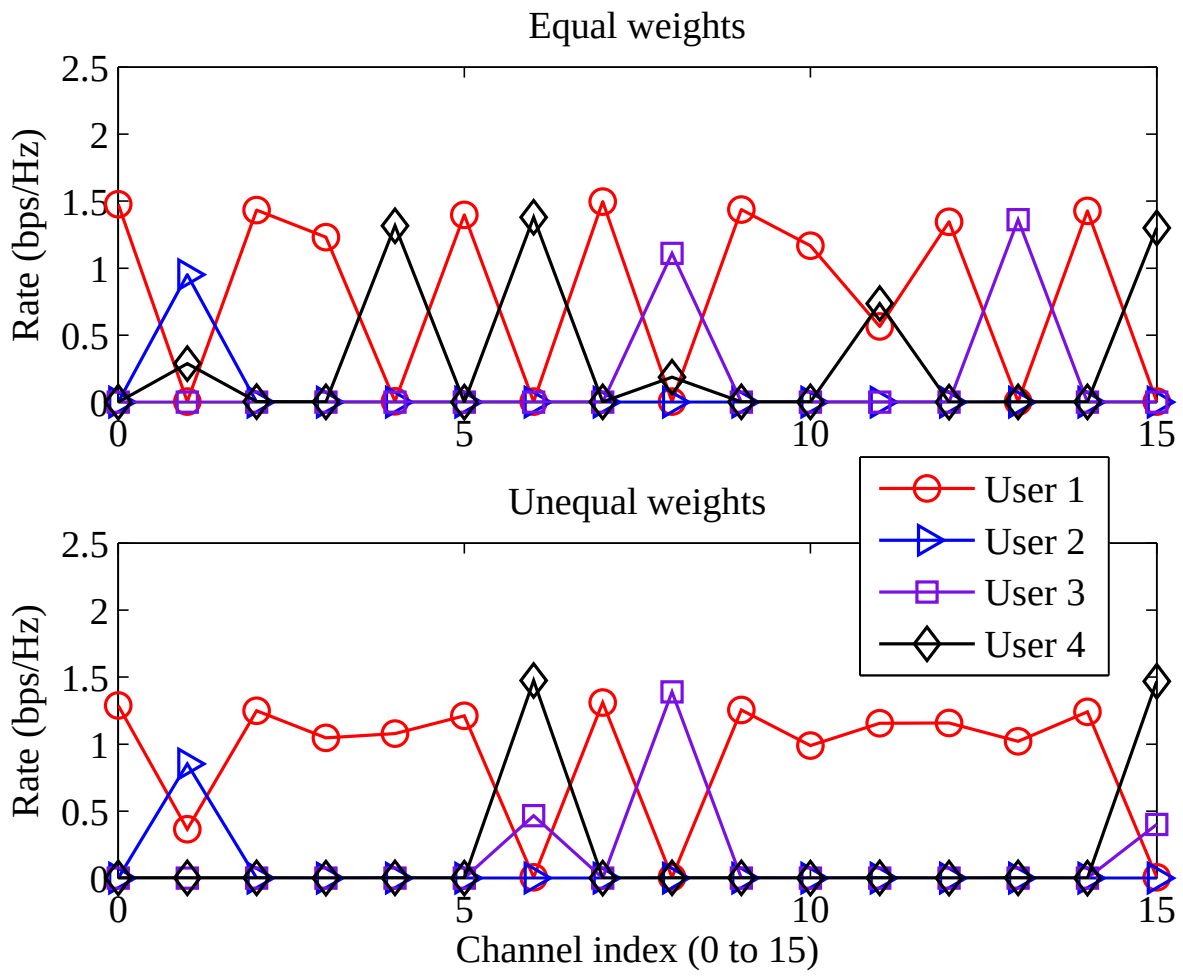

Fig. 4. Optimal rate allocation under Algorithm RPD

Unless otherwise specified, we assume the following parameter setting throughout all simulations. The variables are initialized as $x_{ij} = 1/M$, $p_{ij} = P_i/N$, $\mu_j = 0$, and $\lambda_i = 0.01 \times \max_j(w_i e_{ij})$ for all $i$ and $j$. The update stepsizes in Algorithm RPD are chosen as $k^x_{ij} = 10^{-2}$, $k^\mu_j = 10^{-1} + \varepsilon_j$, and $k^\lambda_i = 10^{-2}$ for all $i$ and $j$. Here, $\varepsilon_j$'s for all channels are chosen to be very small values and diverse from each other, in order to meet the requirement of Corollary 1 as the sufficient condition for system convergence. Each user has a total transmission power constraint $P_i = 2$Watts. Users' channel conditions are randomly generated from the simulator, and users have equal weights $w_i = 1$ for all $i$.

### *A. Optimal Resource Allocation*

We first demonstrate how the power and subchannel are allocated once Algorithm RPD converges to the optimal solution. In order to show the results effectively, we consider a small network with 4 users and 16 subchannels. Larger networks with be simulated in Sections V-B to V-D. The initial values of $\lambda_i$'s are chosen to be $0.1 \times \max_j(w_i e_{ij})$ for all $i$. The self-noise coefficient is $\beta = 0.01$, and there are no maximum SNR constraints (i.e., $s_{ij} = \infty$ for all $i$ and $j$). Fig. 1 shows the channel conditions of 4 users. The horizontal axis is the channel index (from 0 to 15), and the vertical axis is the normalized channel gain $e_{ij}$. It is clear that user 1 has the best channel condition and user 2 has the worst. We simulate Algorithm RPD under two different priority weight settings: (a) *equal weights* where $w_i = 1$ for each user $i$, and (b) *unequal weights* where $w_1 = 2$, $w_2 = w_3 = 1$, and $w_4 = 0.5$.

[5] Every 8 adjacent tones are grouped into one subchannel. This corresponds to the "Band AMC mode" of 802.16 d/e and can help to reduce the feedback overhead. For discussions on various ways of subchannelization, see [7].

Figures 2, 3, and 4 show the optimal channel allocation, power allocation, and rate allocation, respectively. In all three

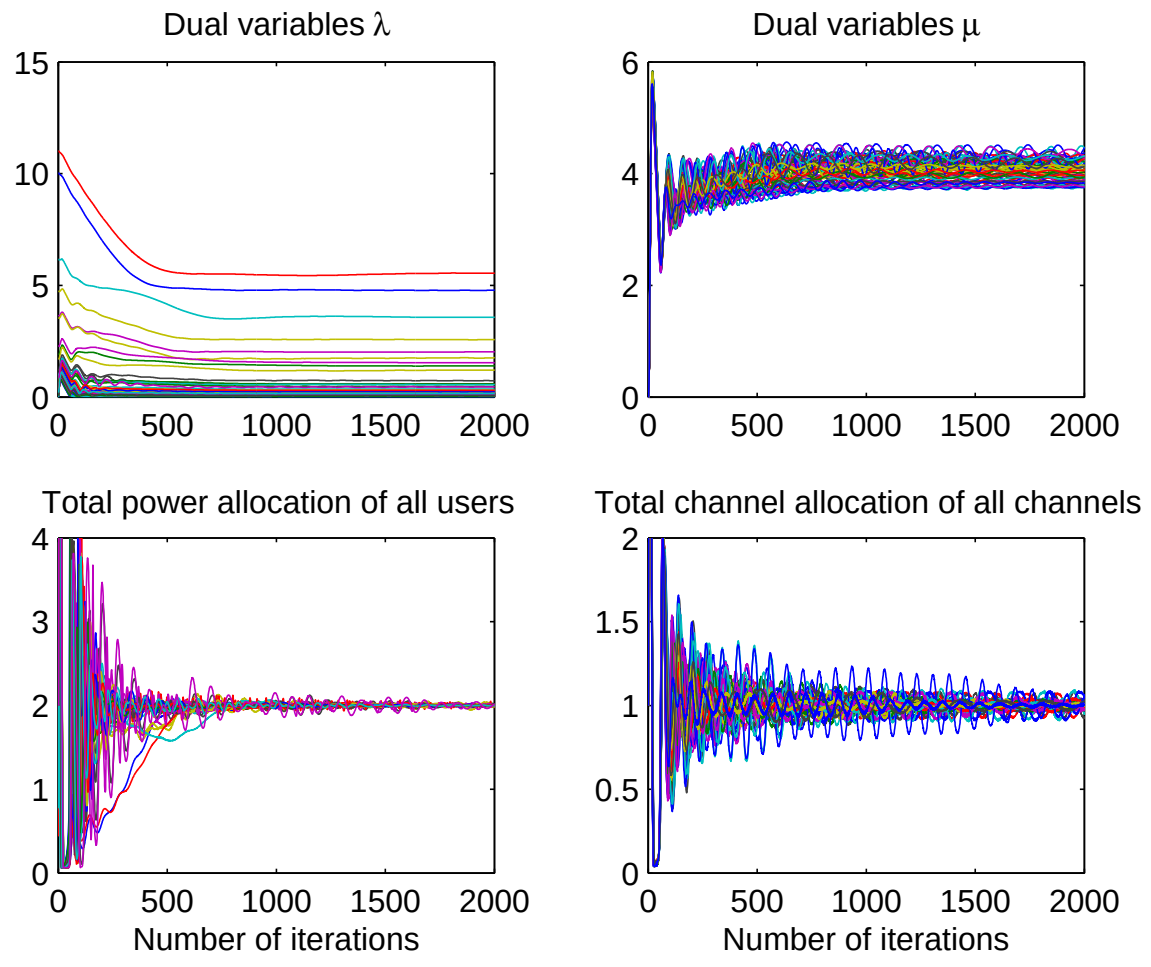

Fig. 5. Case 1 for 40 users: primal and dual variable convergence of Algorithm RPD

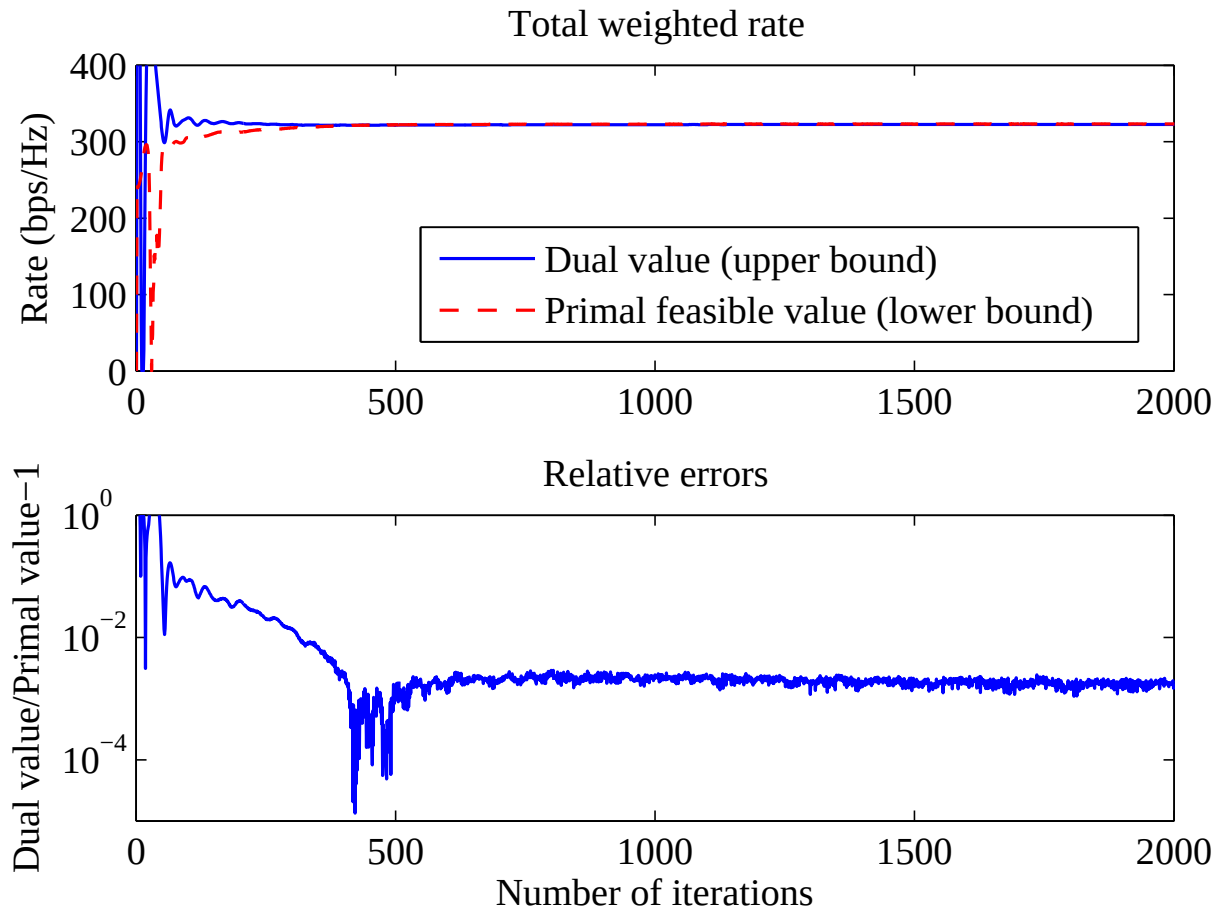

Fig. 6. Case 1 for 40 users: total weighted rate convergence of Algorithm RPD

figures, the top and bottom subfigures correspond to equal weights and unequal weights, respectively. Now consider a user with a good channel condition and a higher weight, e.g., user 1. It is clear that user 1 is allocated with larger fractions of channels (Fig. 2). Since the total power constraint for all users is the same, user 1 actually transmits with less power on average over the subchannels allocated to him (Fig. 3). The net result is that he achieves a better data rate (Fig. 4).

### *B. Algorithm Convergence*

*1) 40 Users:* Next we show the convergence of Algorithm RPD with 40 users and 64 subchannels. Here we assume 4 cases as shown in Table II.

For Case 1, Figure 5 shows the convergence of dual variables (upper two subgraphs, $\lambda_i$ for 40 users and $\mu_j$ for 64 subchannels) and primal variables (lower two subgraphs, $\sum_j p_{ij}$ for 40 users and $\sum_i x_{ij}$ for 64 subchannels).

In Fig. 6, the upper subgraph shows how the dual value and primal feasible value change with iterations. The dual value is an upper bound of the global optimal solution of

TABLE II
FOUR CASES

| Case | 1 | 2 | 3 | 4 |
|---|---|---|---|---|
| Self-noise coefficient $\beta$ | 0 | 0 | 0.01 | 0.01 |
| SNR constraints $s_{ij}$ (dB) | $\infty$ | 20 | $\infty$ | 20 |

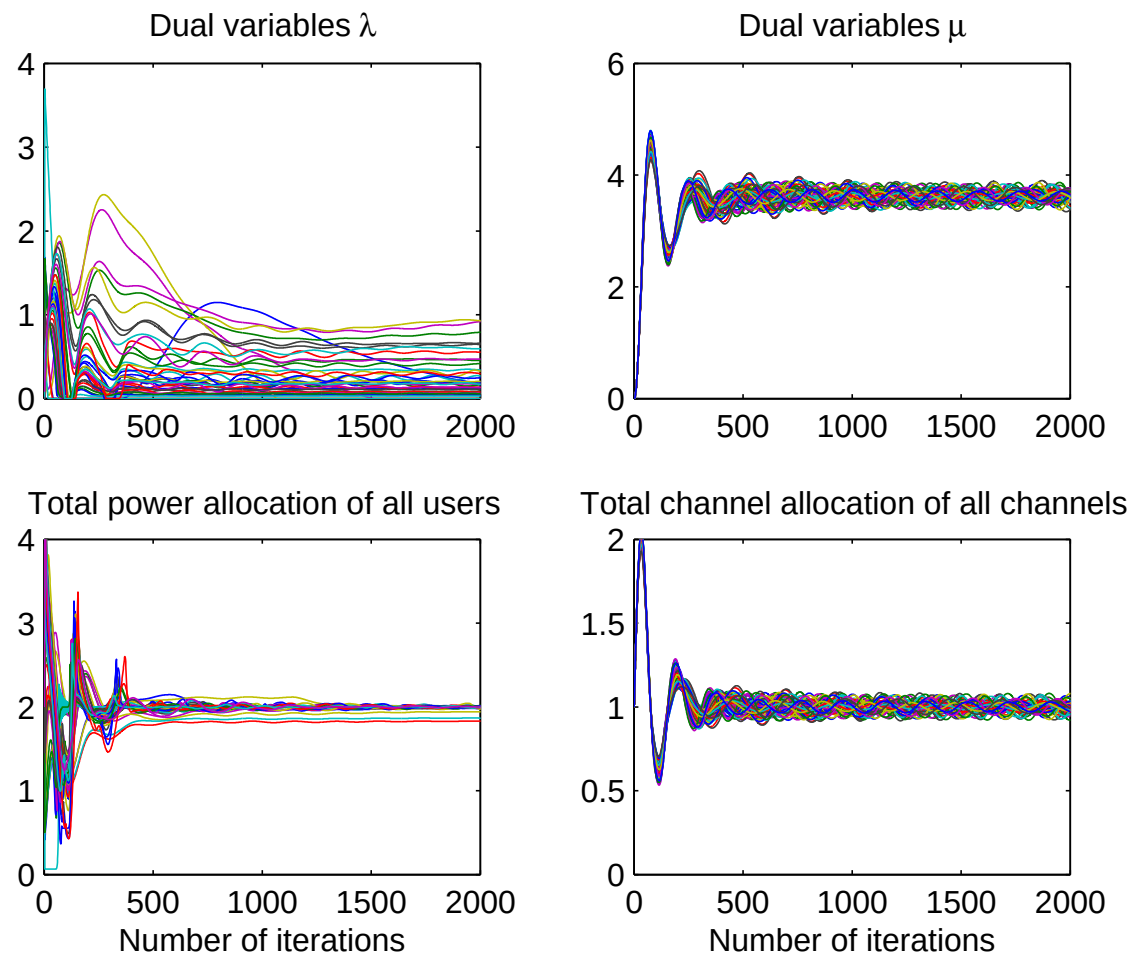

Fig. 7. Case 2 for 40 users: primal and dual variable convergence of Algorithm RPD

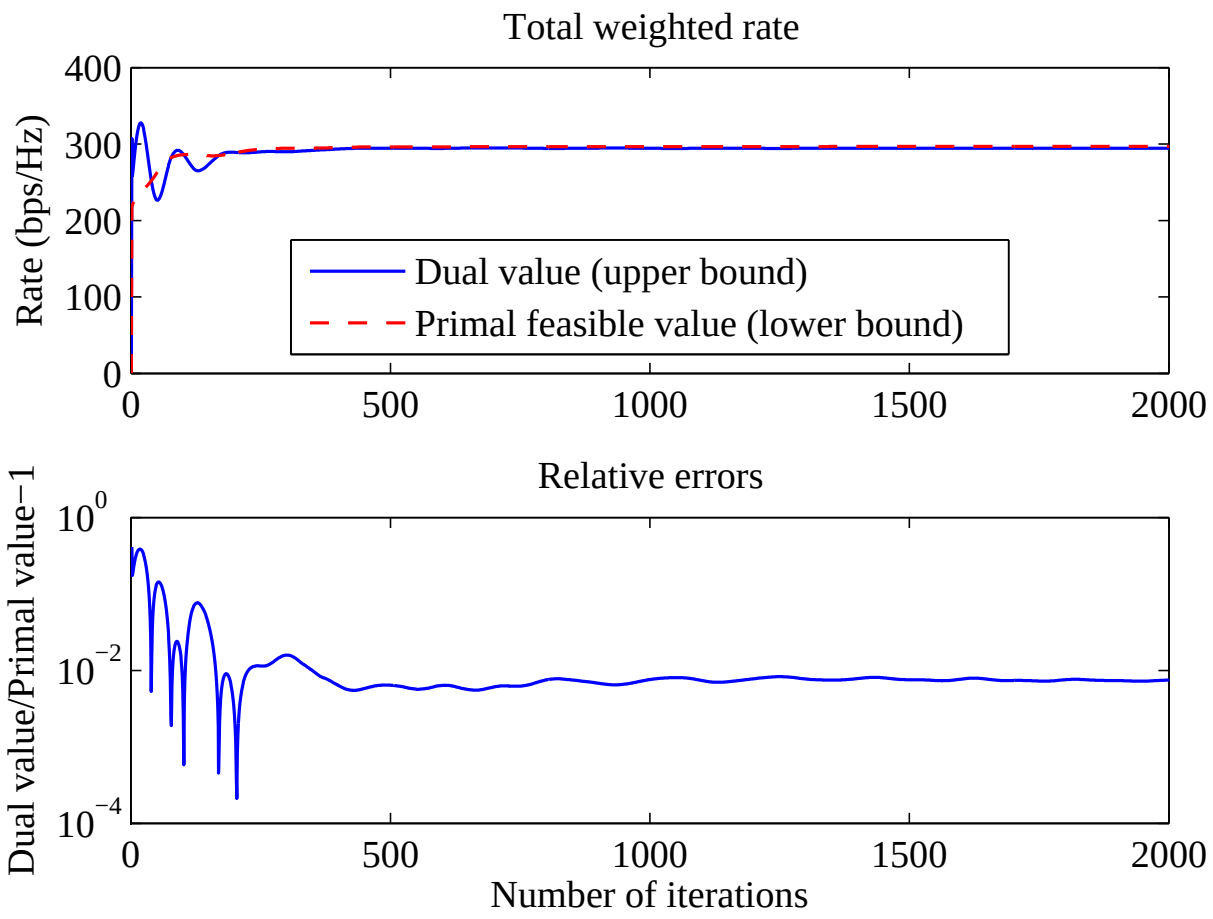

Fig. 8. Case 2 for 40 users: total weighted rate convergence of Algorithm RPD

Problem 2. The primal feasible value is a lower bound and is calculated as follows: given the primal values of $\boldsymbol{p}(t)$ and $\boldsymbol{x}(t)$ at iteration $t$, normalize so that they are feasible and the resources are fully utilized (i.e., $\tilde{p}_{ij}(t) = p_{ij}(t)P_i/(\sum_j p_{ij}(t))$ and $\tilde{x}_{ij}(t) = x_{ij}(t)/(\sum_i x_{ij}(t))$ ), and calculate the achievable rate accordingly. The bottom subfigure shows the relative errors of two curves plotted in the upper subfigure. If we define the stopping criterion to be the relative error less than $5 \times 10^{-3}$, then Algorithm RPD converges in 364 iterations.

We also simulate Cases 2-4 in Table II, and the simulation results are shown in Fig. 7 to Fig. 12.

*2) 20 Users:* Next we show the convergence of Algorithm RPD with 20 users and 64 subchannels. We also consider the 4 cases in Table II. Simulation results for the convergence of rates are plotted in Fig. 13 to Fig. 16.

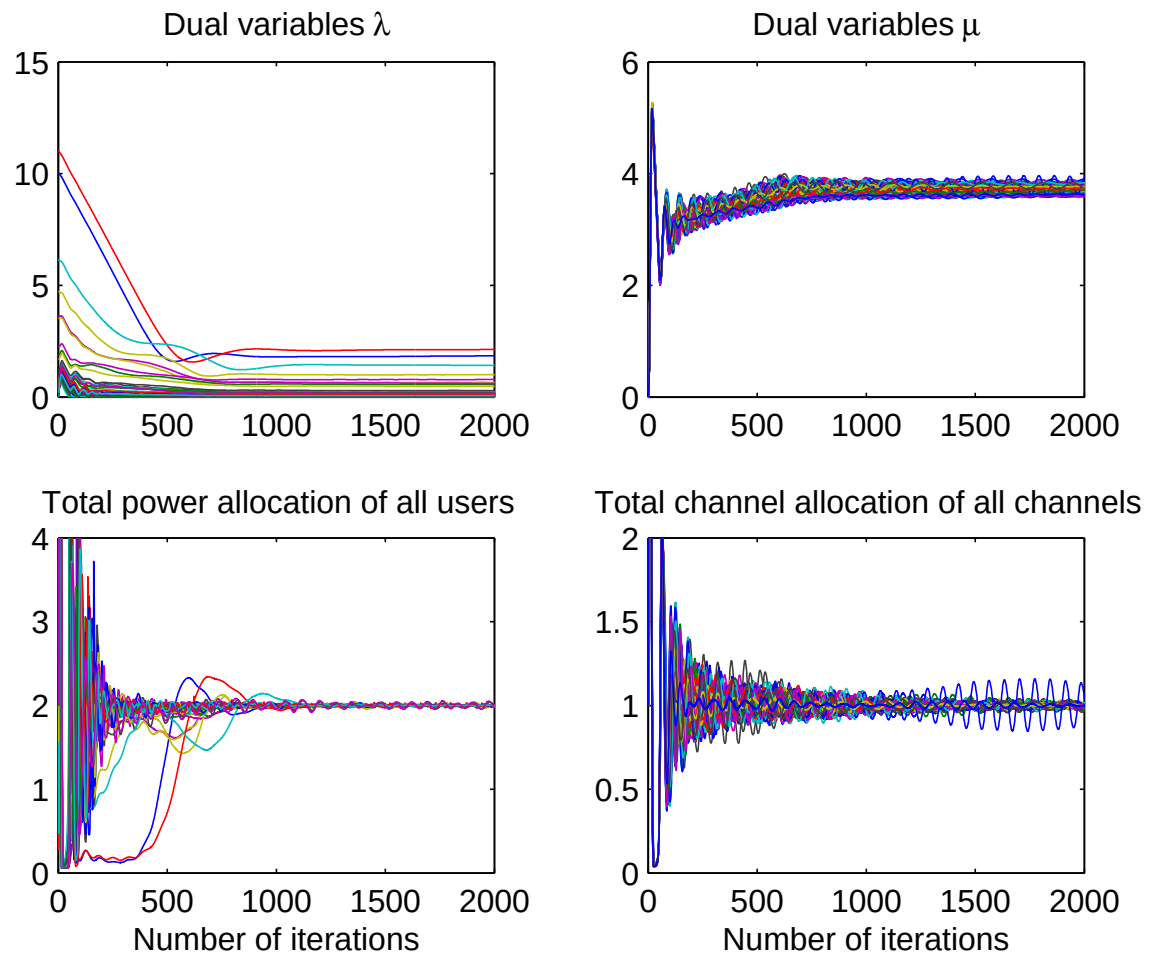

Fig. 9. Case 3 for 40 users: primal and dual variable convergence of Algorithm RPD

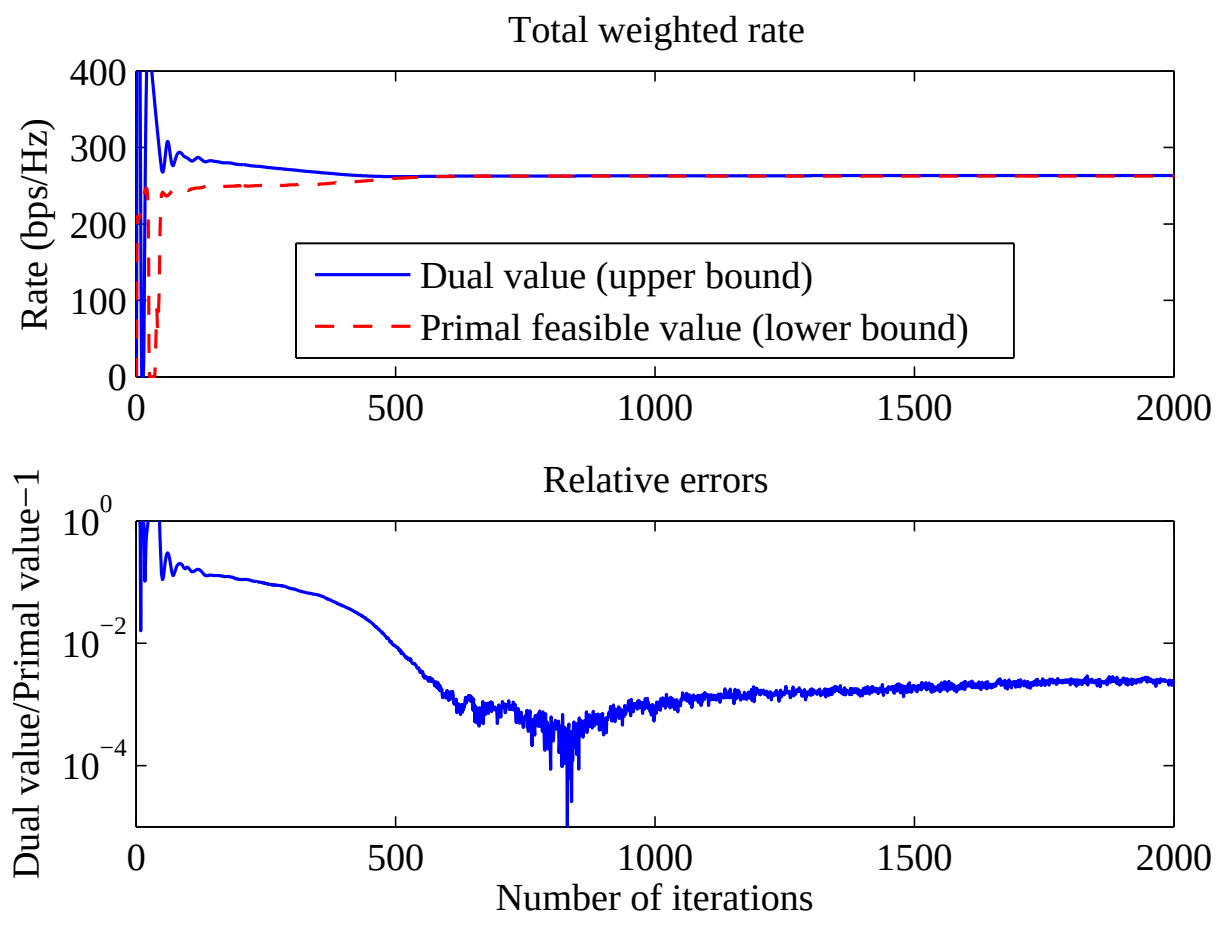

Fig. 10. Case 3 for 40 users: total weighted rate convergence of Algorithm RPD

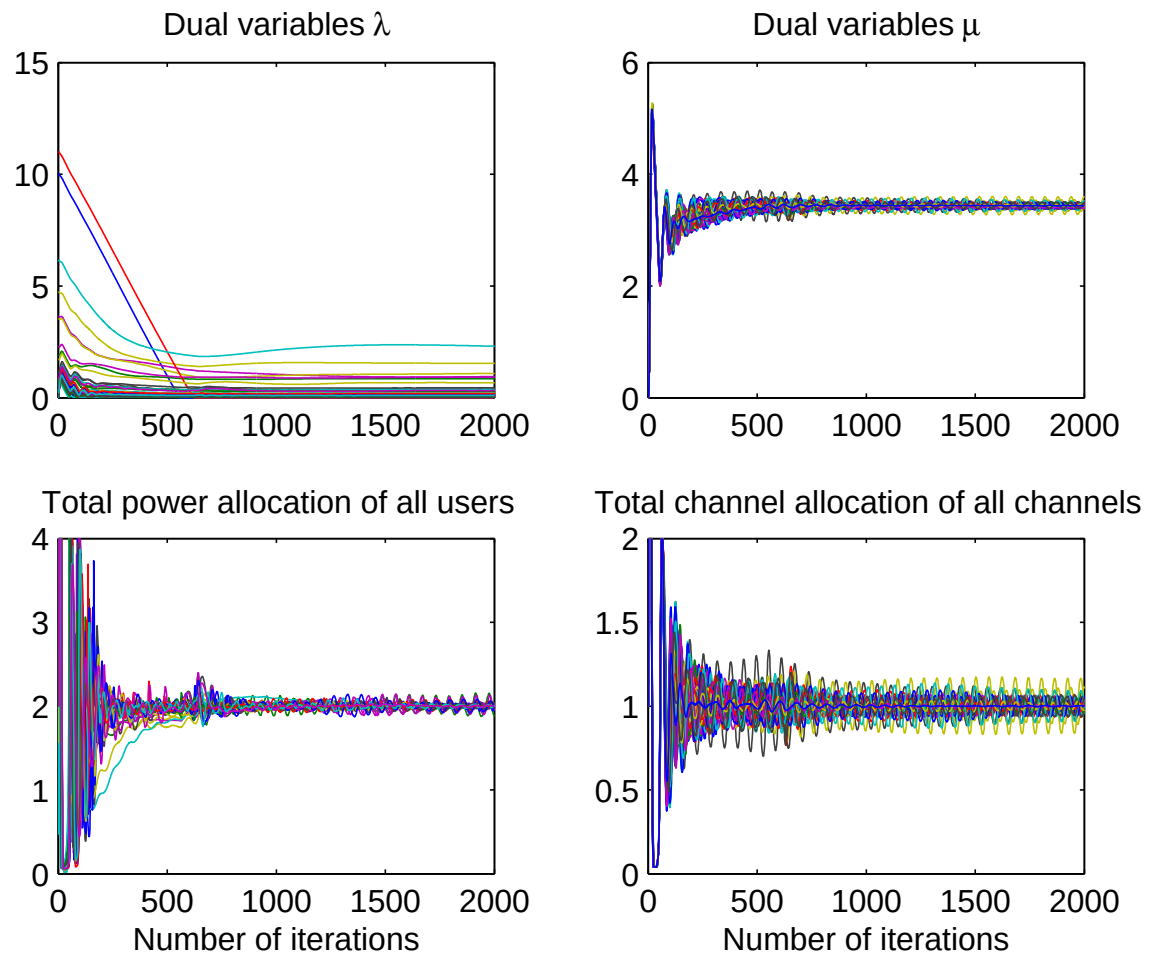

Fig. 11. Case 4 for 40 users: primal and dual variable convergence of Algorithm RPD

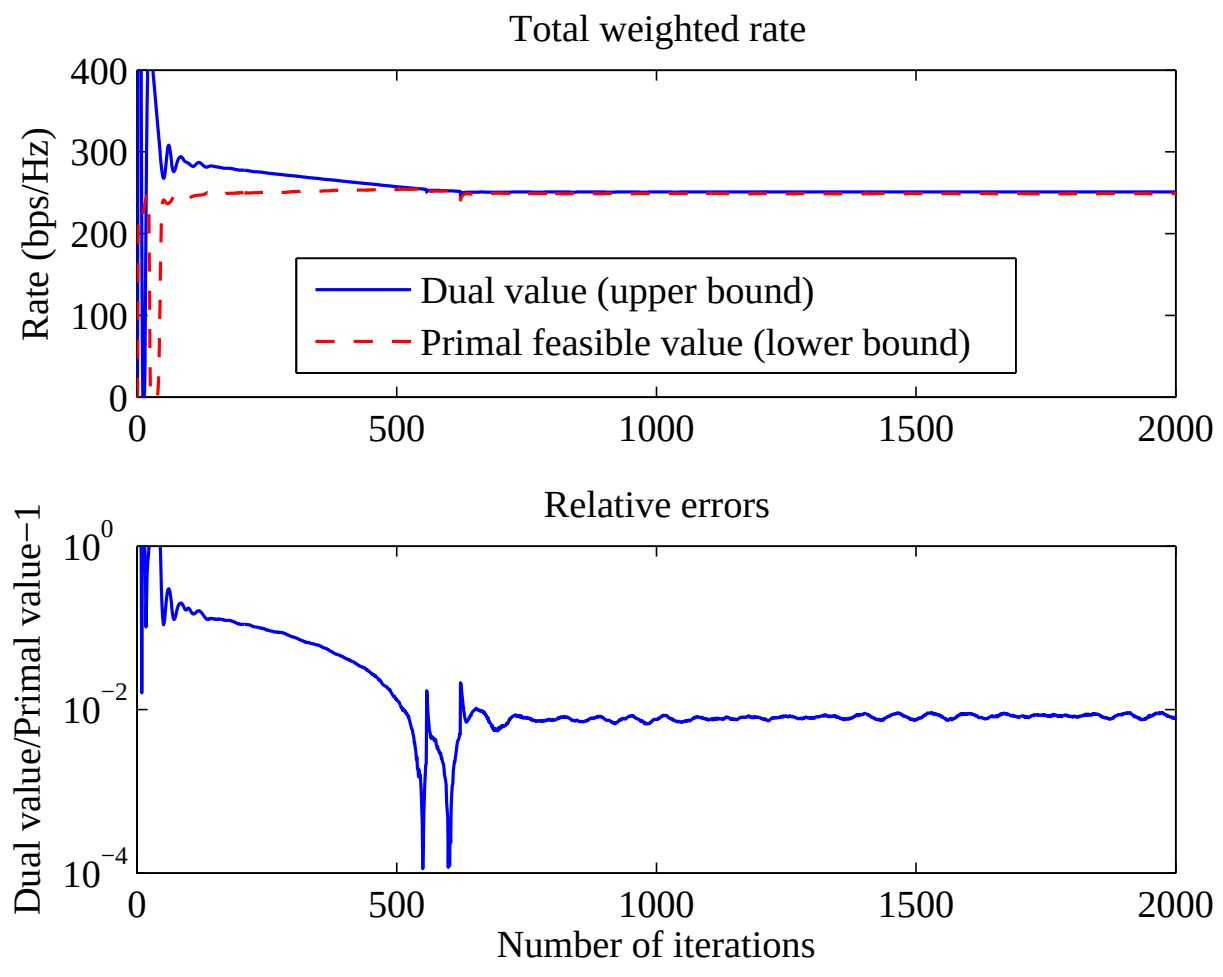

Fig. 12. Case 4 for 40 users: total weighted rate convergence of Algorithm RPD

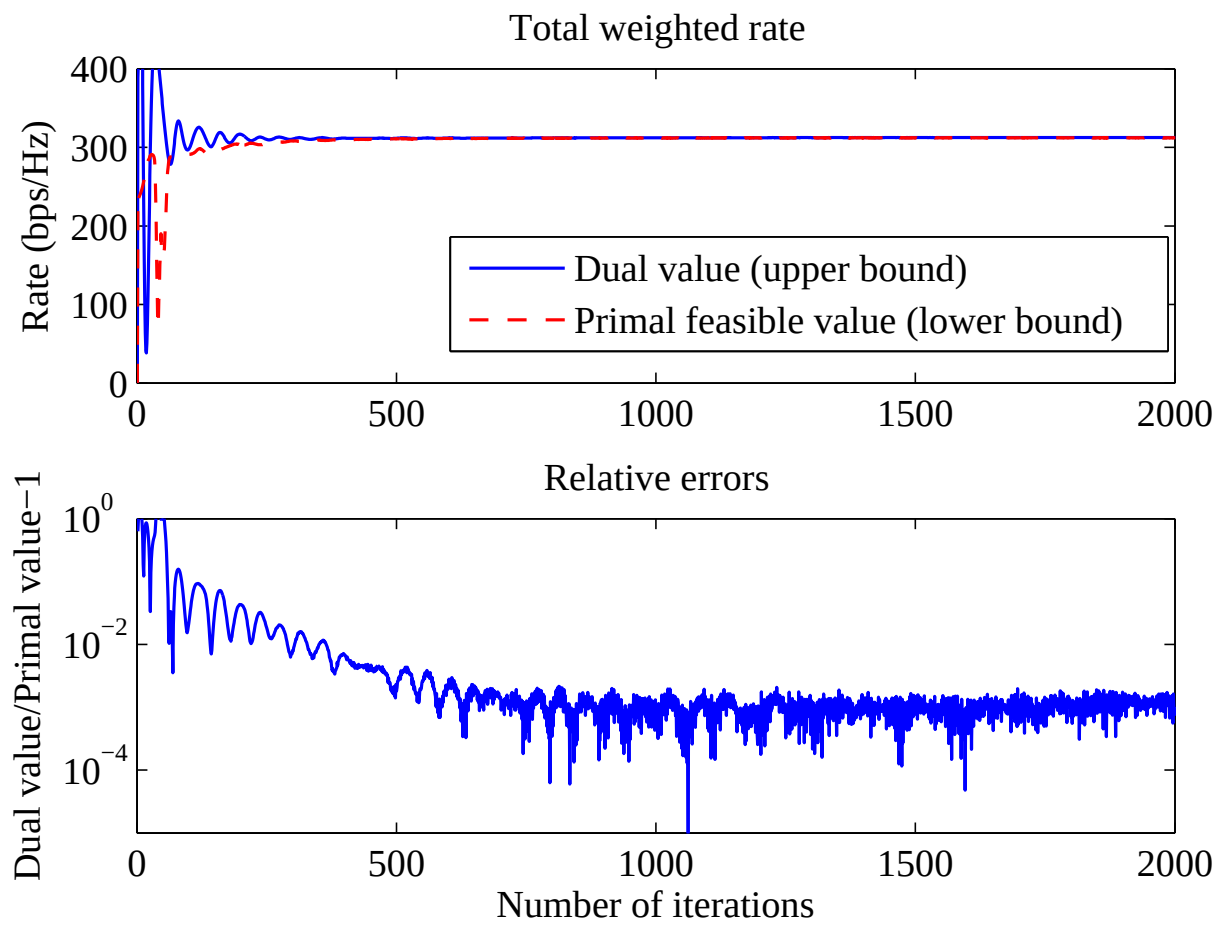

Fig. 13. Case 1 for 20 users: total weighted rate convergence of Algorithm RPD

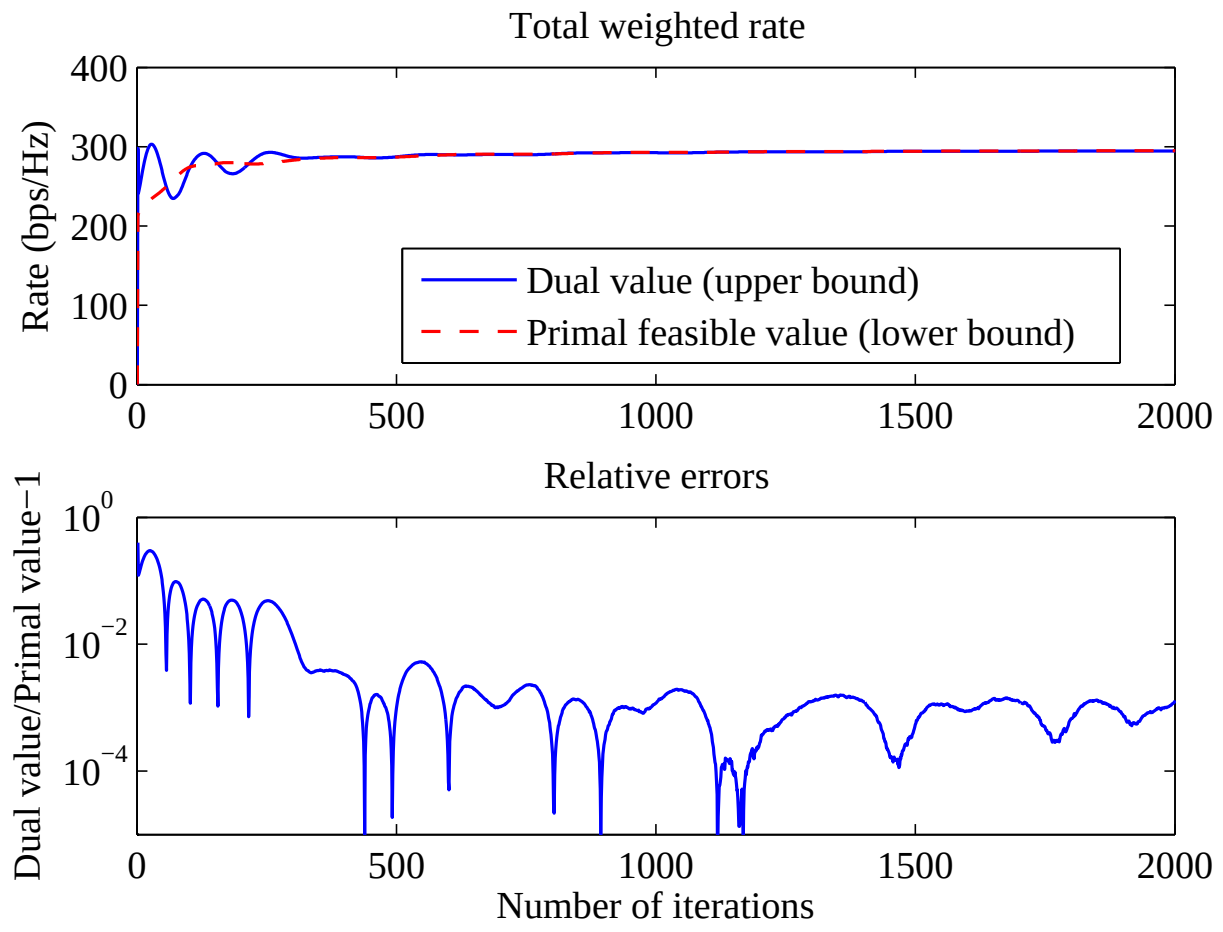

Fig. 14. Case 2 for 20 users: total weighted rate convergence of Algorithm RPD

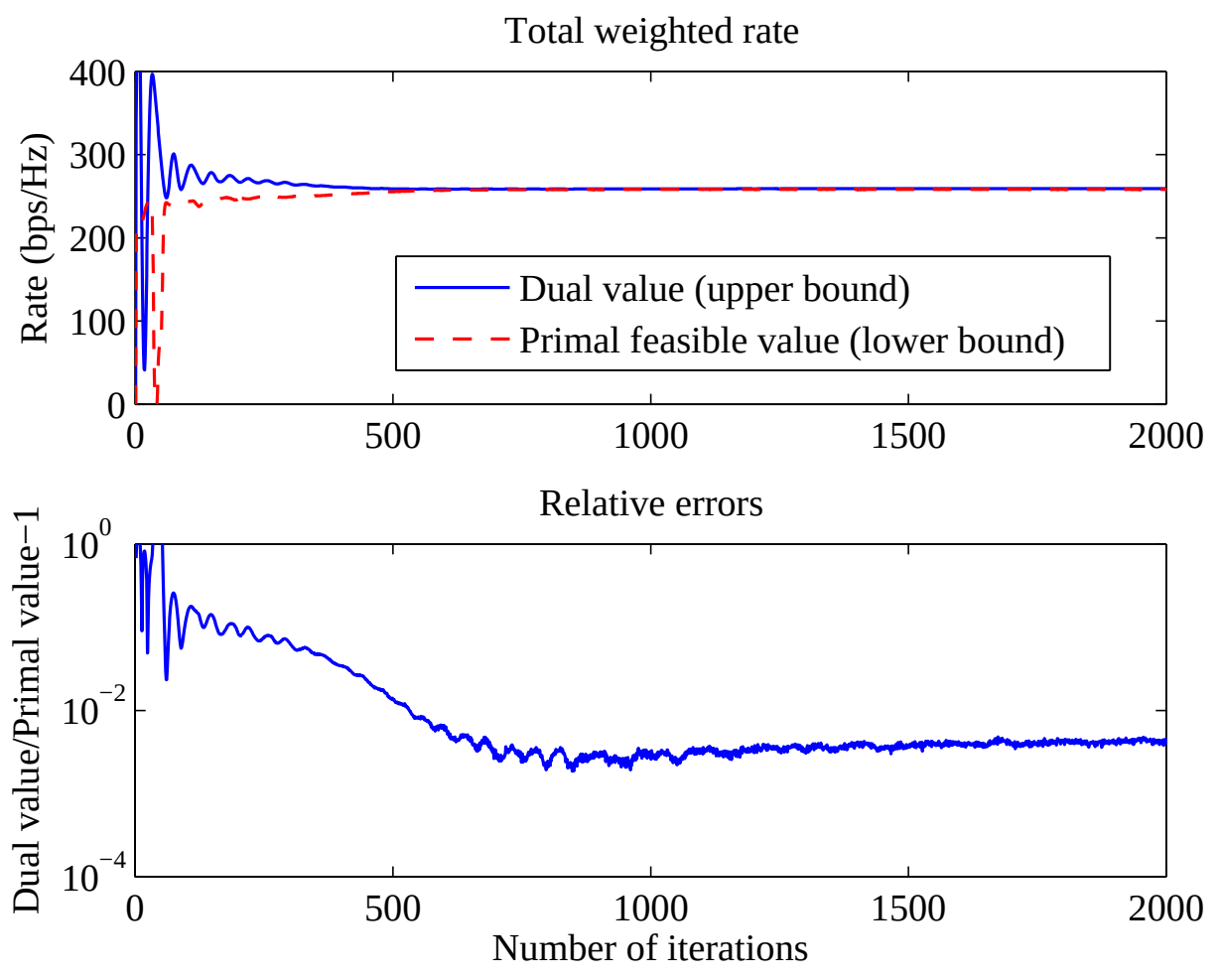

Fig. 15. Case 3 for 20 users: total weighted rate convergence of Algorithm RPD

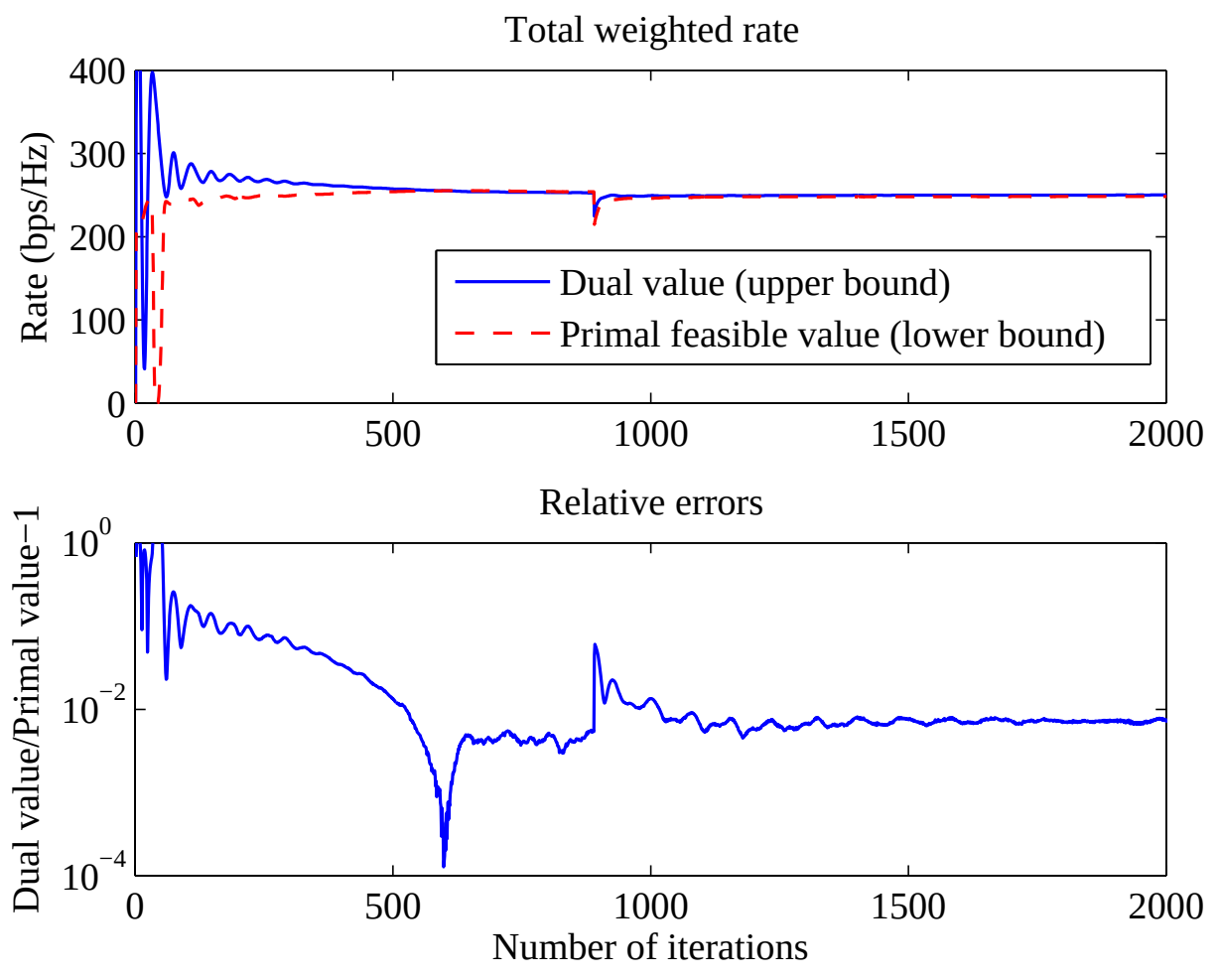

Fig. 16. Case 4 for 20 users: total weighted rate convergence of Algorithm RPD

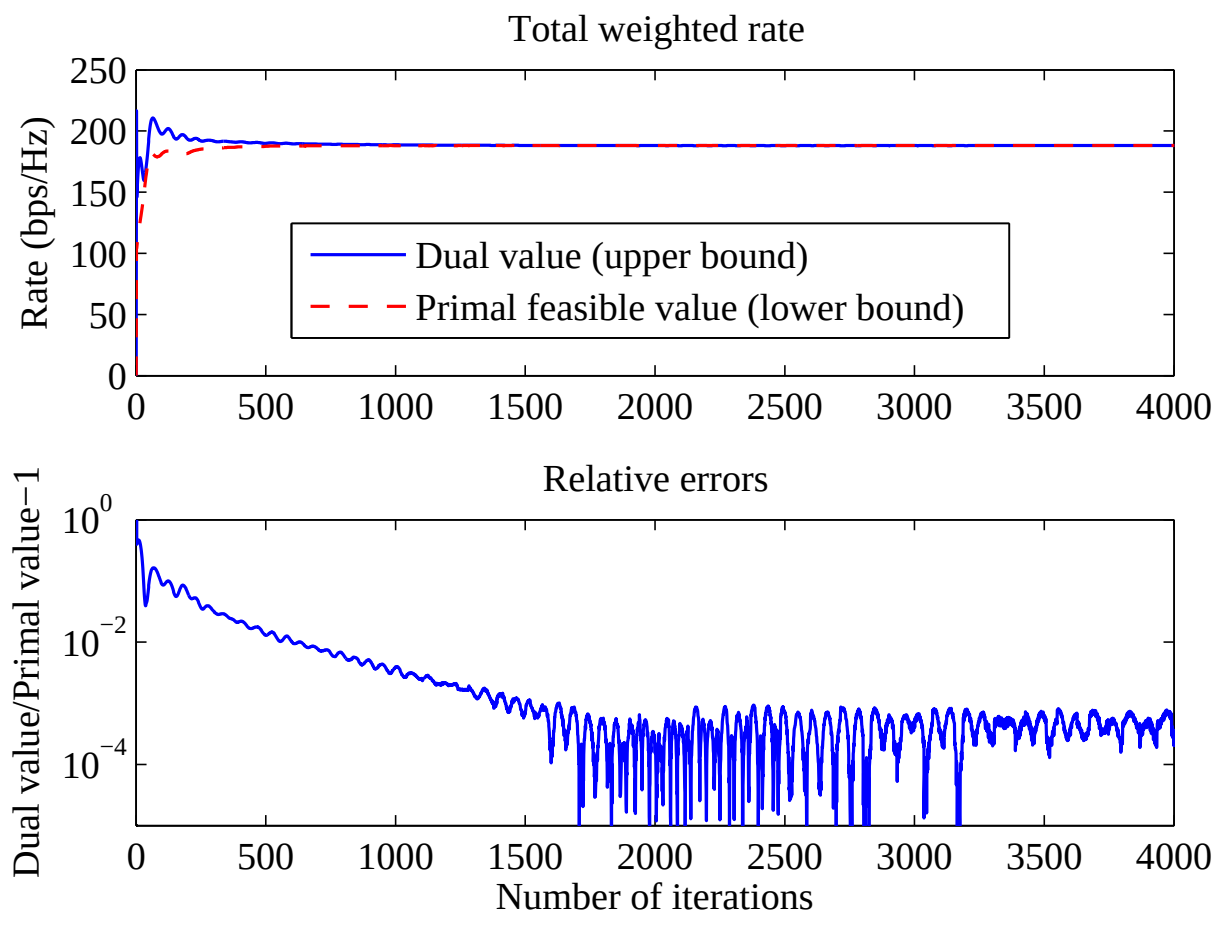

Fig. 17. Cases 1-2 for 4 users: total weighted rate convergence of Algorithm RPD

*3) 4 Users:* Next we show the convergence of Algorithm RPD with 4 users and 64 subchannels. Simulation results of the rate convergence for the 4 cases in Table II are plotted in Fig. 17 and Fig. 18. Here, Cases 1-2 have the same results, as shown in Fig. 17, and Cases 3-4 have the same results, as shown in Fig. 18.

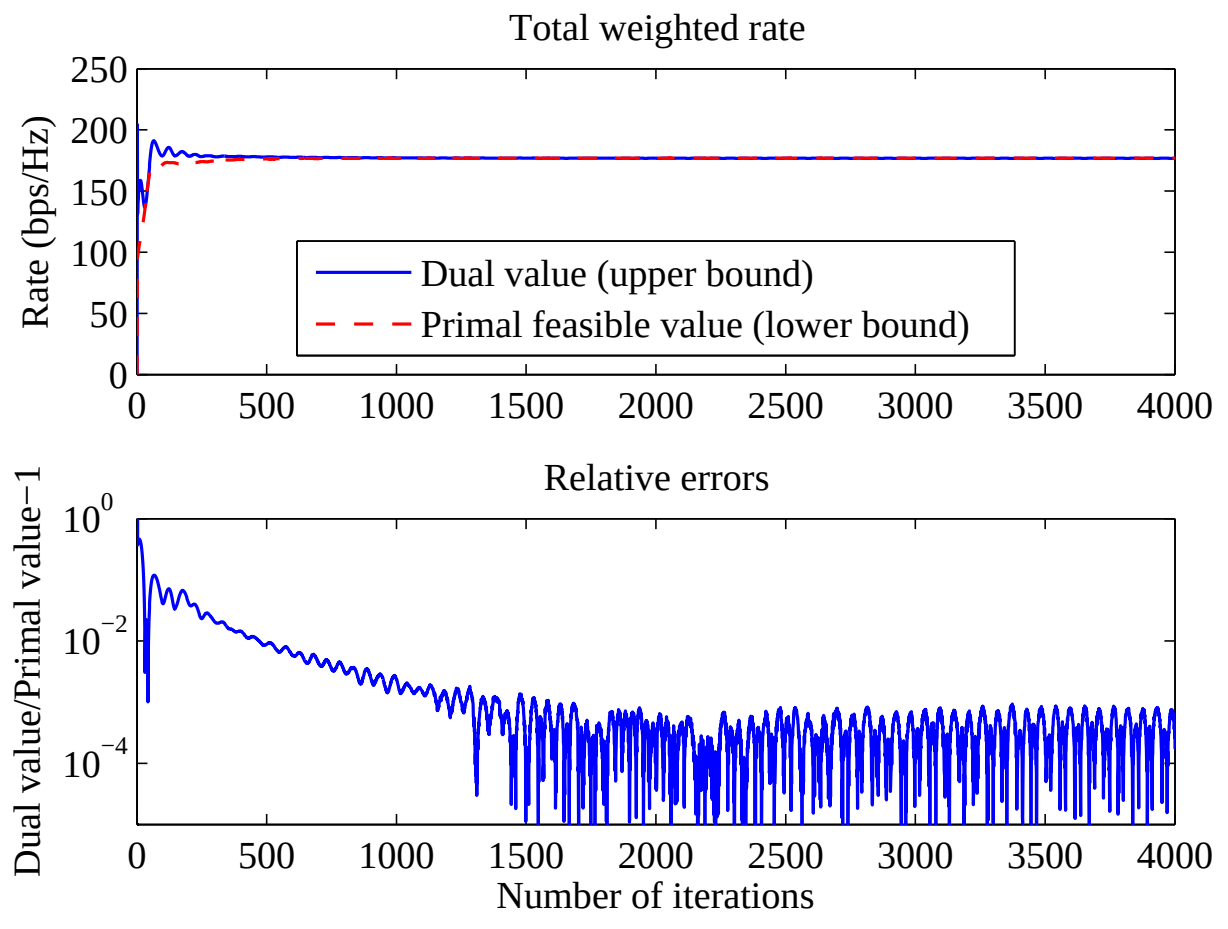

Fig. 18. Cases 3-4 for 4 users: total weighted rate convergence of Algorithm RPD

TABLE III
IMPACT OF SYSTEM PARAMETERS WITH 40 USERS

| Case | 1 | 2 | 3 | 4 |
|---|---|---|---|---|
| Self-noise coefficient $\beta$ | 0 | 0 | 0.01 | 0.01 |
| SNR constraints $s_{ij}$ (dB) | $\infty$ | 20 | $\infty$ | 20 |
| Total weighted rate (Mbps) | 25.12 | 24.51 | 20.47 | 19.96 |
| Convergence time (iterations) | 364 | 355 | 531 | 532 |

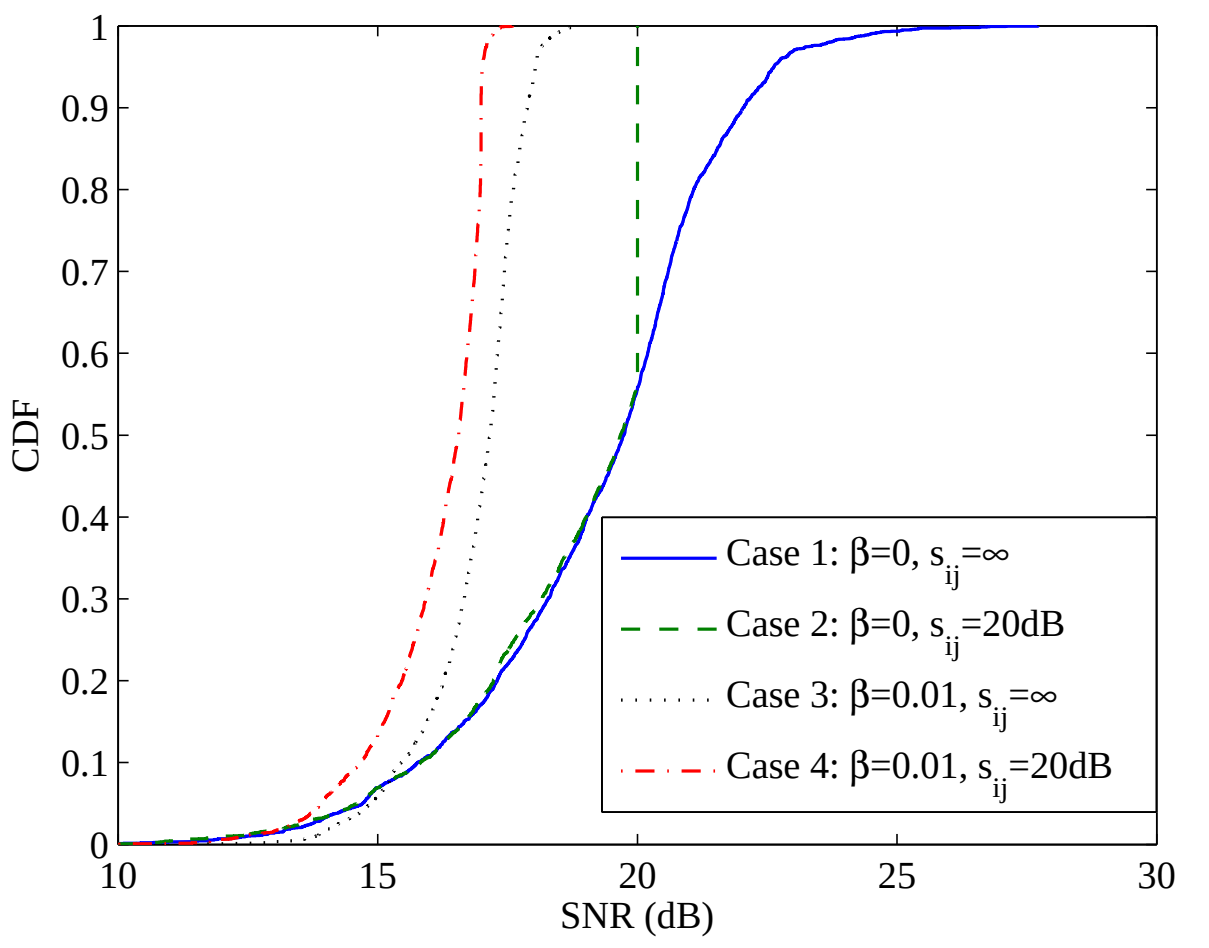

Fig. 19. 40 users: CDF of users' achievable SNRs

## C. Impact of System Parameters

In Table III, we demonstrate how different system parameters affect the convergence and performance of the algorithm. In all cases, we have 40 users and 64 subchannels. Users' channel conditions are the same as in Section V-B1. It is clear from Table III that both positive self-noise and finite maximum SNR constraints reduce the system performance and typically increase the convergence time.

In Fig. 19, we plot the Cumulative Distribution Function

TABLE IV
IMPACT OF SYSTEM PARAMETERS WITH 20 USERS

| Case | 1 | 2 | 3 | 4 |
|---|---|---|---|---|
| Self-noise coefficient $\beta$ | 0 | 0 | 0.01 | 0.01 |
| SNR constraints $s_{ij}$ (dB) | $\infty$ | 20 | $\infty$ | 20 |
| Total weighted rate (Mbps) | 24.27 | 22.31 | 20.20 | 20.01 |
| Convergence time (iterations) | 375 | 319 | 610 | 551 |

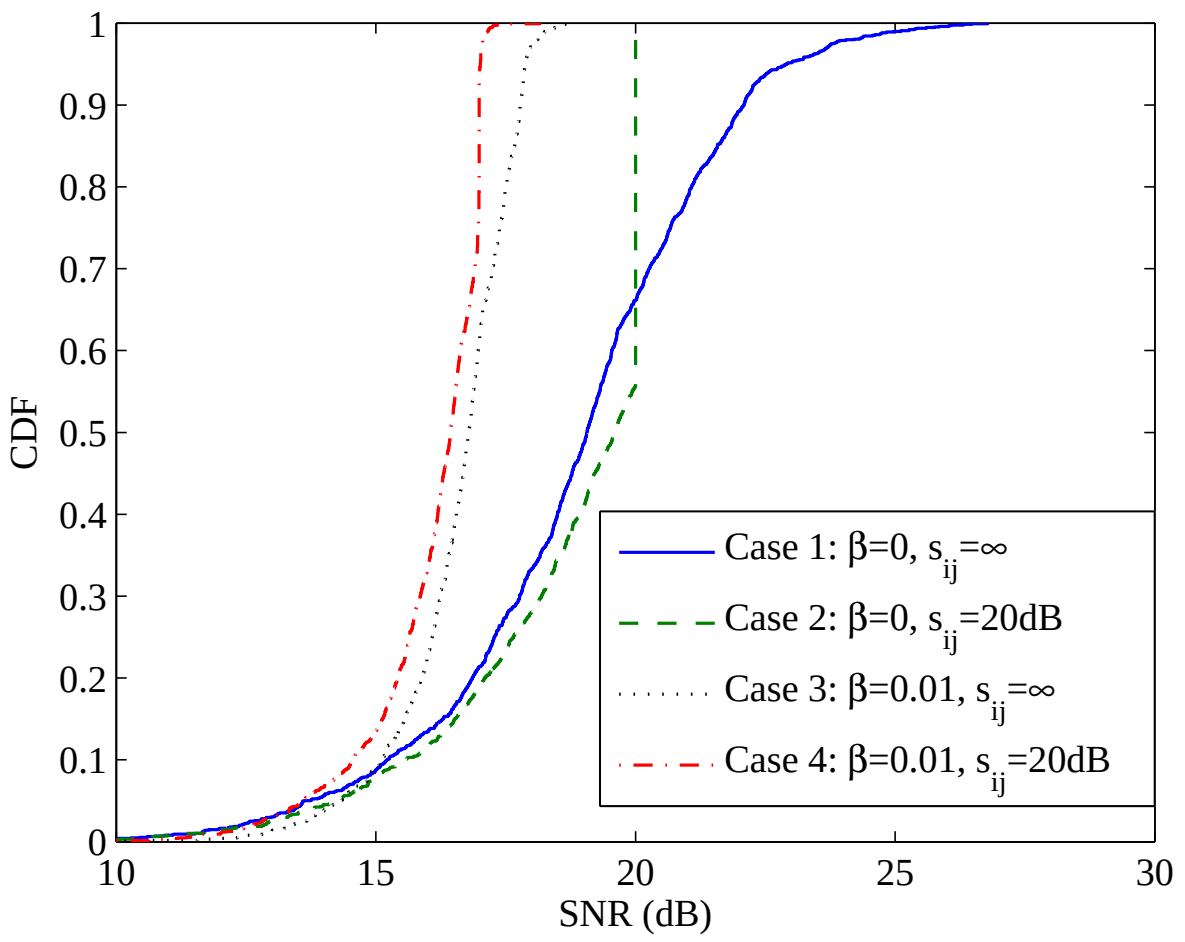

Fig. 20. 20 users: CDF of users' achievable SNRs

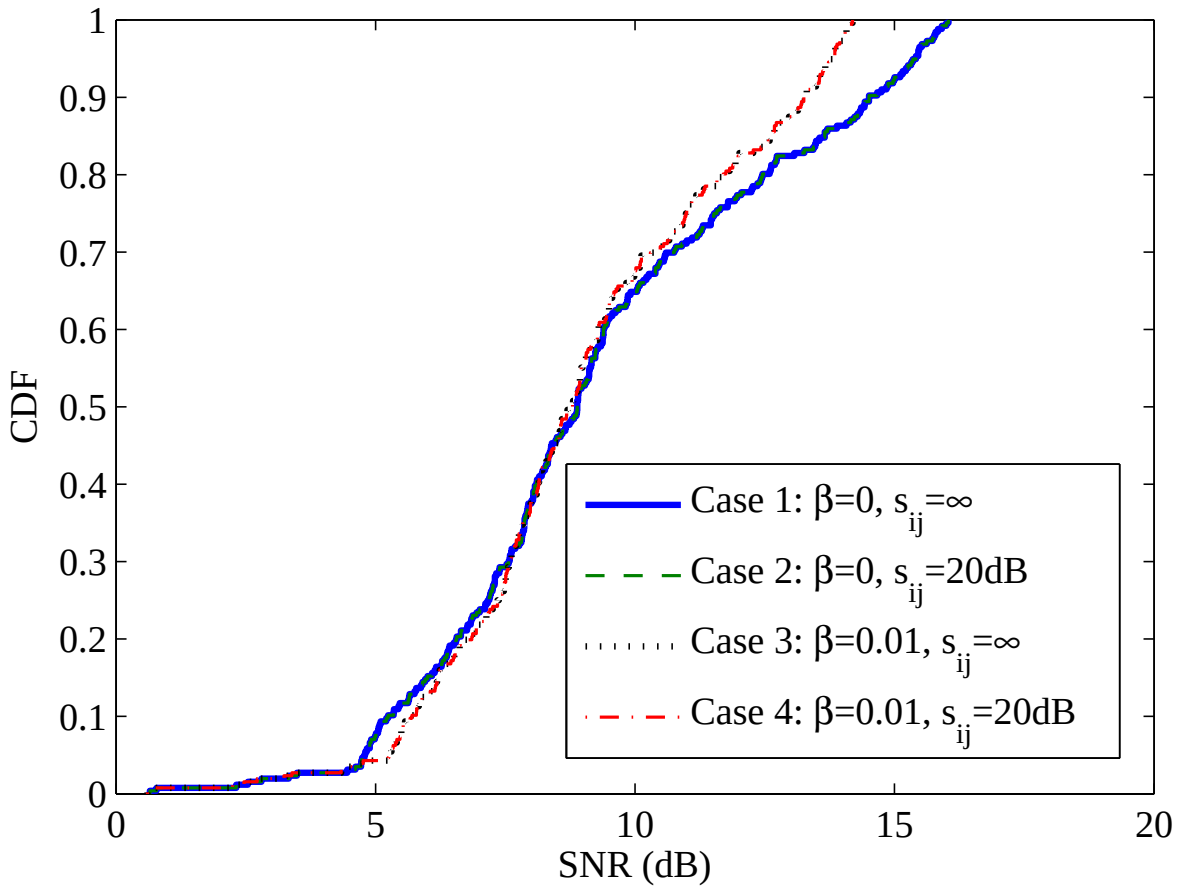

Fig. 21. 4 users: CDF of users' achievable SNRs

(CDF) of users' allocated SNRs $\left(p_{ij}e_{ij}/\left(x_{ij}+\beta p_{ij}e_{ij}\right)\right)$ of all four cases after Algorithm RPD converges. In Case 2, the SNRs are "truncated" at 20dB due to $s_{ij} = 20$dB. In Case 3, the SNRs are also no larger than 20dB since they are effectively upper bounded by $1/\beta$. However, the effects of self-noise and maximum SNR constraints are apparently different. Case 4 shows a further reduction in SNR due to both factors.

We also give results for the case of 20 users under the same conditions as Section V-B2. The simulation results are shown in Table IV and Fig. 20.

Similarly, we show results for the case of 4 users under the same conditions as Section V-B3 in Table V and Fig. 21.

TABLE V
IMPACT OF SYSTEM PARAMETERS WITH 4 USERS

| Case | 1 | 2 | 3 | 4 |
|---|---|---|---|---|
| Self-noise coefficient $\beta$ | 0 | 0 | 0.01 | 0.01 |
| SNR constraints $s_{ij}$ (dB) | $\infty$ | 20 | $\infty$ | 20 |
| Total weighted rate (Mbps) | 14.76 | 14.76 | 13.87 | 13.87 |
| Convergence time (iterations) | 858 | 858 | 644 | 644 |

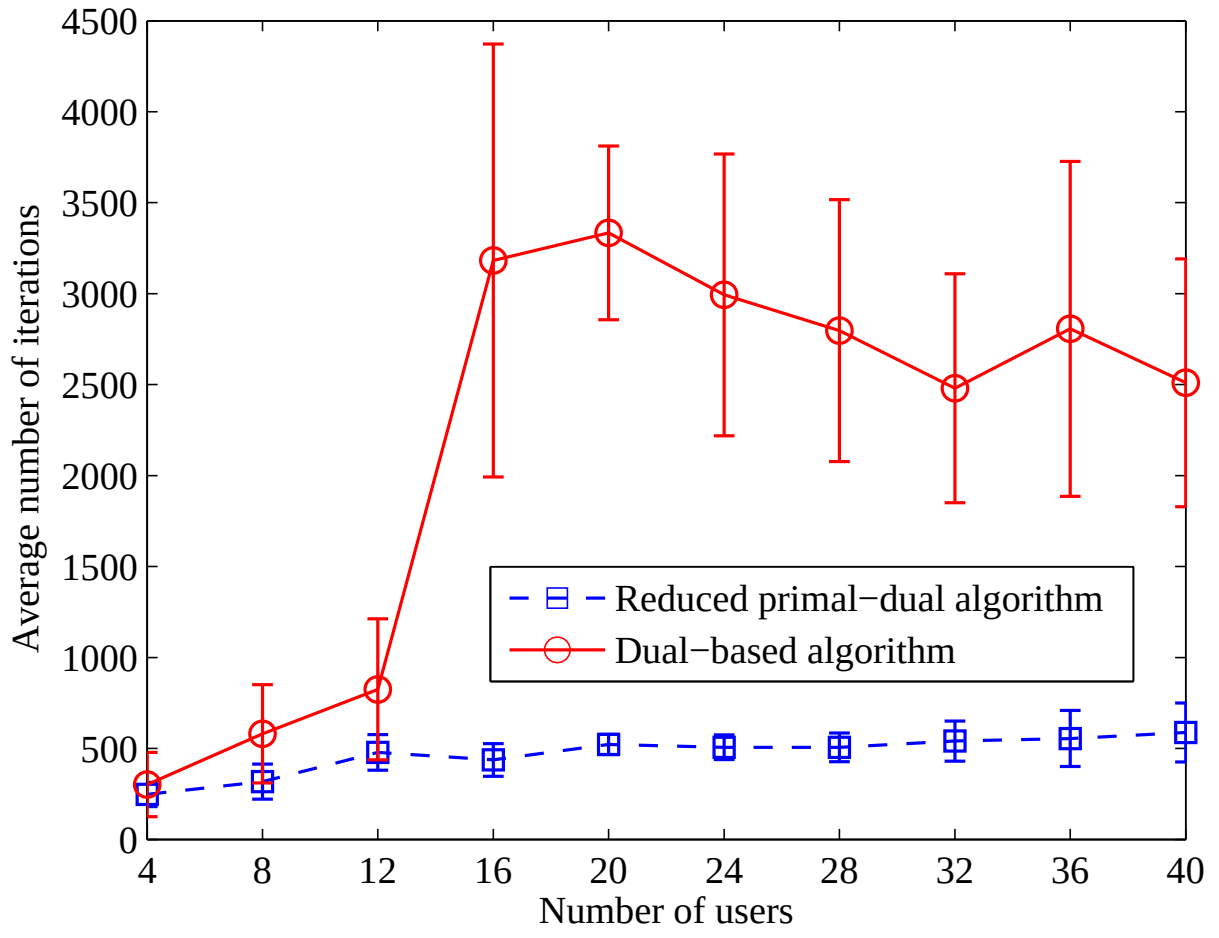

Fig. 22. Average number of iterations and standard deviation of reduced primal-dual algorithm and dual-based algorithm

### D. Comparison with the Dual-based Algorithm

In Fig. 22, we compare the convergence speed of Algorithm RPD with the dual-based centralized optimal solution proposed in [15]. The self-noise coefficient is $\beta = 0.1$, and the maximum SNR coefficient is $s_{ij} = 20$dB for all $i$ and $j$. We vary the number of users from 4 to 40. For a fixed user population size, we randomly generated 10 sets of different weights and channel conditions. We plot both the average and the standard deviation (i.e., error bar) of the number of iterations for both algorithms as the number of users changes. In all cases, the reduced primal-dual algorithm converges with less number of iterations and a much smaller variance.[6]

## VI. Conclusion and Future Work

We presented the first distributed optimal primal-dual resource allocation algorithm for uplink OFDM systems. The key features of the proposed algorithm include: (a) incorporating practical OFDM system constraints such as self-noise and maximum SNR constraints, (b) reduced primal-dual algorithm which eliminates unnecessary variable updates, (c) distributed implementation by the end users and base station, (d) simple local updates with limited message passing, (e) global convergence despite of the existence of multiple global optimal solutions, and (f) fast convergence compared with the state-of-art centralized optimal algorithm. Currently we are extending this work in several directions, including proving the theoretical convergence speed of the algorithm.

[6]We note that the time needed for each iteration is significantly less for the centralized algorithm, which does not require frequent message passing between users and the base station.

## Appendix

### A. Proof of Theorem 1

Let $(\boldsymbol{x}^*, \boldsymbol{p}^*, \boldsymbol{\mu}^*, \boldsymbol{\lambda}^*)$ be one point satisfying the KKT condition. Motivated by the Lyapunov function used in [5], we consider the following La Salle function

$$V(x,\mu,\lambda) = \sum_{i,j} \frac{1}{k^x_{ij}} \int_0^{x_{ij}} (\xi - x^*_{ij}) d\xi + \sum_j \frac{1}{k^\mu_j} \int_0^{\mu_j} (\xi - \mu^*_j) d\xi + \sum_i \frac{1}{k^\lambda_i} \int_0^{\lambda_i} (\xi - \lambda^*_i) d\xi.$$

It is straightforward to verify that $V$ is semi-positive definite. Its Lie derivative over an invariant set $\{(x,\mu,\lambda) |\ x \geq 0,\ \mu \geq 0,\ \lambda \geq 0\}$ is given by

$$\begin{aligned}
\dot{V} &= \sum_{i,j} \frac{\partial V}{\partial x_{ij}} \dot{x}_{ij} + \sum_j \frac{\partial V}{\partial \mu_j} \dot{\mu}_j + \sum_i \frac{\partial V}{\partial \lambda_i} \dot{\lambda}_i \\
&\leq \sum_{i,j} (x_{ij} - x^*_{ij})(f_{ij} - \mu_j) \\
&\quad + \sum_j (\mu_j - \mu^*_j)(\sum_i x_{ij} - 1) + \sum_i (\lambda_i - \lambda^*_i)(\sum_j p_{ij} - P_i) \\
&= \sum_{i,j} (\lambda_i - g_{ij})(p_{ij} - p^*_{ij}) \\
&\quad + \sum_j (\mu_j - \mu^*_j)(\sum_i x^*_{ij} - 1) + \sum_i (\lambda_i - \lambda^*_i)(\sum_j p^*_{ij} - P_i) \\
&\quad + \sum_{i,j} ([f^*_{ij}, g^*_{ij}] - [\mu^*_j, \lambda^*_i]) \begin{bmatrix} x_{ij} - x^*_{ij} \\ p_{ij} - p^*_{ij} \end{bmatrix} \\
&\quad + \sum_{i,j} ([f_{ij}, g_{ij}] - [f^*_{ij}, g^*_{ij}]) \begin{bmatrix} x_{ij} - x^*_{ij} \\ p_{ij} - p^*_{ij} \end{bmatrix}, \qquad (29)
\end{aligned}$$

where $f^*_{ij} = f_{ij}(x^*_{ij}, p^*_{ij})$ and $g^*_{ij} = g_{ij}(x^*_{ij}, p^*_{ij})$.

Now let's look at the expression in (29). The first sum is non-positive, since $(\lambda_i - g_{ij})(p_{ij} - p^*_{ij}) = 0$ when $p_{ij} > 0$ and $(\lambda_i - g_{ij})(p_{ij} - p^*_{ij}) \leq 0$ when $p_{ij} = 0$ according to (19). The second sum is also non-positive, since $\sum_j x^*_{ij} \leq 1$ and $\mu^*_j = 0$ when $\sum_j x^*_{ij} < 1$ according to (9). In the same way, the third sum is also non-positive according to (10).

As to the fourth sum, it has $f^*_{ij} \leq u^*_j$ according to (12), and the inequality holds only when $x^*_{ij} = 0$. Similarly, $g^*_{ij} \leq \lambda^*_i$ according to (13), and the inequality holds only when $p^*_{ij} = 0$. When $x^*_{ij} \neq 0$, we have $p^*_{ij} \neq 0$ and

$$[f^*_{ij}, g^*_{ij}] = [u^*_j, \lambda^*_i].$$

Hence, the fourth sum of (29) is also non-positive.

For the fifth sum, first we notice $[f_{ij}, g_{ij}]^T$ is the gradient of the following function

$$H_{ij}(x_{ij}, p_{ij}) = w_i(x_{ij} + \epsilon_{ij}) \log\left(1 + \frac{p_{ij} e_{ij}}{x_{ij} + \beta p_{ij} e_{ij} + \epsilon_{ij}}\right).$$

By concavity of the function $H_{ij}$, we have

$$H_{ij}(x_{ij}, p_{ij}) \leq H_{ij}(x^*_{ij}, p^*_{ij}) + \nabla H_{ij}(x^*_{ij}, p^*_{ij}) \begin{bmatrix} x_{ij} - x^*_{ij} \\ p_{ij} - p^*_{ij} \end{bmatrix}, \quad (30)$$

$$H_{ij}(x^*_{ij}, p^*_{ij}) \leq H_{ij}(x_{ij}, p_{ij}) + \nabla H_{ij}(x_{ij}, p_{ij}) \begin{bmatrix} x^*_{ij} - x_{ij} \\ p^*_{ij} - p_{ij} \end{bmatrix}. \quad (31)$$

Consequently, the inner product of the gradient difference and the variable difference is non-positive, i.e.

$$([f_{ij}, g_{ij}] - [f^*_{ij}, g^*_{ij}]) \begin{bmatrix} x_{ij} - x^*_{ij} \\ p_{ij} - p^*_{ij} \end{bmatrix} \leq 0;$$

hence, the fifth sum is also non-positive.

As such, we have $\dot{V} \leq 0$. According to La Salle principle [20], trajectories of the system in (21) to (23) converge to the set $V_0 = \left\{(x,\mu,\lambda) : \dot{V} = 0\right\}$ globally and asymptotically. It is clear that at any optimal solution of Problem 2 we will have $\dot{V} = 0$, i.e., such solution is in set $V_0$.

Over set $V_0$, the five sums in (29) must all equal to zero so as to ensure $\dot{V} = 0$. Then we have the following observations:

- $\mu_j$ is nonzero only if $\sum_i x^*_{ij} = 1$ (according to (9) and the fact that the second sum in (29) equals to zero);
- $\left([f_{ij}, g_{ij}] - [f^*_{ij}, g^*_{ij}]\right) \begin{bmatrix} x_{ij} - x^*_{ij} \\ p_{ij} - p^*_{ij} \end{bmatrix} = 0$ (since the fifth sum in (29) equals to zero).

Combining the second observation with (30) and (31), we further know that $\forall (x,p,\mu,\lambda) \in V_0$,

$$H_{ij}(x_{ij}, p_{ij}) = H_{ij}(x^*_{ij}, p^*_{ij}) + \nabla H_{ij}(x^*_{ij}, p^*_{ij}) \begin{bmatrix} x_{ij} - x^*_{ij} \\ p_{ij} - p^*_{ij} \end{bmatrix}.$$

Taking the derivative of both sides, we have $\forall (x,\mu,\lambda) \in V_0$,

$$\begin{bmatrix} f_{ij}(x_{ij}, p_{ij}) \\ g_{ij}(x_{ij}, p_{ij}) \end{bmatrix} = \nabla H_{ij}(x^*_{ij}, p^*_{ij}) = \text{constant}.$$

From $g_{ij}(x_{ij}, p_{ij}) = g_{ij}(x^*_{ij}, p^*_{ij})$, we can derive

$$\begin{aligned}
&\left(1 + \beta e_{ij} \frac{p_{ij}}{x_{ij} + \epsilon_{ij}}\right)\left(1 + (\beta+1) e_{ij} \frac{p_{ij}}{x_{ij} + \epsilon_{ij}}\right) \\
=&\left(1 + \beta e_{ij} \frac{p^*_{ij}}{x^*_{ij} + \epsilon_{ij}}\right)\left(1 + (\beta+1) e_{ij} \frac{p^*_{ij}}{x^*_{ij} + \epsilon_{ij}}\right). \qquad (32)
\end{aligned}$$

It can be verified that the function in both sides in (32), i.e. $\left(1 + \beta e_{ij} y\right)\left(1 + (\beta+1) e_{ij} y\right)$, is a monotonically increasing function of $y$. Therefore, Eq. (32) implies $\frac{p_{ij}}{x_{ij}+\epsilon_{ij}} = \frac{p^*_{ij}}{x^*_{ij}+\epsilon_{ij}}$.

Suppose there exists an optimal solution $\{x^*_{ij}, p^*_{ij}\}$ with $\sum_j p^*_{ij} < P_i$. For the user $i$ on one certain subchannel $l$, it must have $\hat{p}_{il} = P_i - \sum_{j \neq l} p^*_{ij} > p^*_{il}$, which indicates that there always exists a new allocated value $\hat{p}_{il}$ larger than the optimal $p^*_{il}$. By substituting this $\hat{p}_{il}$ into (7) and keeping all rest $p^*_{ij}$ $(\forall i, \forall j \neq l)$ unchanged, the value of the objective function in Problem 2 is increased. That contradicts with the assumption that $\{x^*_{ij}, p^*_{ij}\}$ is optimal. Therefore, for any optimal solution $\{x^*_{ij}, p^*_{ij}\}$, we must have $\sum_j p^*_{ij} = P_i$.

Considering the above result $p_{ij} = (x_{ij} + \epsilon_{ij}) \frac{p^*_{ij}}{x^*_{ij}+\epsilon_{ij}}$ in set $V_0$, it has at least one nonzero $p_{ij}$. Consequently, with the fact $\lambda_i = g_{ij}(x_{ij}, p_{ij})$ according to (19) and $g_{ij}(x_{ij}, p_{ij}) = g^*_{ij}$ (which is always positive seen from $g_{ij}$'s definition) discussed above, we get $\lambda_i = \lambda^*_i > 0$, which implies that $\lambda_i$ remains constant over set $V_0$.

### B. Proof of Lemma 1

From (27), we have

$$A_2 B\boldsymbol{x} = P - A_2 B\boldsymbol{\epsilon} = \text{constant}.$$

Result 1 can be derived by taking derivatives (with respect to time) on both sides of the above equation.

For result 2, it can be verified that the characteristic function of the linear system (25)-(26) is a product of positive diagonal matrix and a skew-symmetric matrix,

$$\begin{bmatrix} 0 & -K^x A_1^T \\ K^\mu A_1 & 0 \end{bmatrix} = \begin{bmatrix} K^x & 0 \\ 0 & K^\mu \end{bmatrix} \begin{bmatrix} 0 & -A_1^T \\ A_1 & 0 \end{bmatrix}$$

Since all eigenvalues of a skew-symmetric matrix are either zero or purely imaginary, the eigenvalues of the characteristic function are also imaginary. Hence trajectories of the linear system do not converge.

### C. Proof of Theorem 2

By linear system theory, $\boldsymbol{\mu}$ is completely observable from the constant $A_2 B\boldsymbol{x}$ if and only if the complete observability (28) holds [21]. Then the invariant set $V_0$ contains only the equilibria of the linear system in (25) to (27), which are the global optimal solutions of Problem 2. Consequently, all trajectories of Algorithm RPD converge globally and asymptotically to the global optimal solutions.

### D. Proof of Corollary 1

Let us choose matrix $K^x$ to be $kI$ where $k$ is a positive constant. By direct computation, we get $A_2 B K^x A_1^T = kA_2 B A_1^T$.

First consider the case where $N \leq M$, it can be observed that $A_2 B K^x A_1^T$ (the upper part of the matrix in (28)) has rank $N$, because $A_2 B A_1^T$ has $N$ linearly independent columns under such condition.

Now consider the case where $M < N$, in which case the rank of matrix $kA_2 B A_1^T$ will be less than $N$. In this case we need to look at matrix $kK^\mu A_1 A_1^T - \sigma I$ (the bottom part of the matrix in (28)). Consider the diagonal matrix $K^\mu A_1 K^x A_1^T = kK^\mu A_1 A_1^T$, which can be written as

$$kK^\mu \begin{bmatrix} M & \cdots & 0 \\ \vdots & \ddots & \vdots \\ 0 & \cdots & M \end{bmatrix}.$$

If all diagonal elements of matrix $K^\mu$ take different values, the rank of matrix $kK^\mu A_1 A_1^T - \sigma I$ would be $N - 1$, because the eigenvalue $\sigma$ is one of the diagonal elements of the diagonal matrix $kK^\mu A_1 A_1^T$. Combining with $A_2 B K^x A_1^T$ which has full row rank $M \geq 1$, the combined matrix in (28) satisfies the observability condition in Theorem 2. To sum up, when $K^x = kI$ and all update stepsizes $k_j^\mu$ $(j = 1, \ldots, N)$ take different values, the observability of the linear system is guaranteed.

## References

[1] R. Agrawal, and V. Subramanian, "Optimality of Certain Channel Aware Scheduling Policies," *Proc. of 2002 Allerton Conference*, Oct. 2002.

[2] A. L. Stolyar, "On the asymptotic optimality of the gradient scheduling algorithm for multiuser throughput allocation," *Operations Research*, vol. 53, No. 1, pp. 12–25, 2005.

[3] H. Kushner, and P. Whiting, "Asymptotic properties of proportional-fair sharing algorithms," in *Proc. 40th Annual Allerton Conference on Communication, Control, and Computing*, Oct. 2002.

[4] T. Voice, "Stability of Congestion Control Algorithms with Multi-Path Routing and Linear Stochastic Modelling of Congestion Control," PhD Thesis, University of Cambridge, May, 2006.

[5] M. Chen, and J. Huang, "Optimal Resource Allocation for OFDM Uplink Communication: A Primal-Dual Approach," *CISS conference*, 2008.

[6] M. Chen, M. Ponec, S. Sengupta, J. Li, and P. A. Chou, "Utility Maximization in Peer-to-peer Systems," *ACM SIGMETRICS*, 2008.

[7] J. Huang, V. G. Subramanian, R. Agrawal, and R. Berry, "Downlink scheduling and resource allocation for OFDM systems," under revision of *IEEE Trans. on Wireless*, 2008.

[8] Z. Han, Z. Ji, and K. Liu, "Fair Multiuser Channel Allocation for OFDMA Networks Using Nash Bargaining Solutions and Coalitions," *IEEE Trans. Comm.*, vol. 53, no. 8, pp. 1366–1376, 2005.

[9] S. Pfletschinger, G. Muenz, and J. Speidel, "Efficient subcarrier allocation for multiple access in OFDM systems," in *7th International OFDM Workshop*, 2002.

[10] Y. Ma, "Constrained Rate-Maximization Scheduling for Uplink OFDMA," in *Proc. IEEE MILCOM*, pp. 1–7, Oct. 2007.

[11] K. Kim, Y. Han, and S. Kim, "Joint subcarrier and power allocation in uplink OFDMA systems," *IEEE Comms. Letters*, vol. 9, no. 6, pp. 526–528, 2005.

[12] C. Ng, and C. Sung, "Low complexity subcarrier and power allocation for utility maximization in uplink OFDMA systems," *IEEE Trans. Wireless Comm.*, vol. 7, no. 5 Part 1, pp. 1667–1675, 2008.

[13] K. Kwon, Y. Han, and S. Kim, "Efficient Subcarrier and Power Allocation Algorithm in OFDMA Uplink System," *IEICE Trans. Comm.*, vol. 90, no. 2, pp. 368–371, 2007.

[14] L. Gao, and S. Cui, "Efficient subcarrier, power, and rate allocation with fairness consideration for OFDMA uplink," *IEEE Trans. Wireless Comm.*, vol. 7, no. 5 Part 1, pp. 1507–1511, 2008.

[15] J. Huang, V. Subramanian, R. Berry, and R. Agrawal, "Joint Scheduling and Resource Allocation in Uplink OFDM Systems," *Proc. of Asilomar Conference*, Pacific Grove, CA, Nov. 2007. Journal version submitted to *IEEE Journal on Selected Areas in Communications*, 2008.

[16] H. Jin, R. Laroia, and T. Richardson, "Superposition by position," *IEEE Information Theory Workshop*, March 2006.

[17] J. Lee, H. Lou, and D. Toumpakaris, "Analysis of Phase Noise Effects on Time-Direction Differential OFDM Receivers," *IEEE GLOBECOM*, 2005.

[18] W. Yu, R. Lui, and R. Cendrillon, "Dual optimization methods for multiuser orthogonal frequency division multiplex systems," in *Proceedings of IEEE Globecom*, vol. 1, 2004, pp. 225–229.

[19] X. Lin and N. B. Shroff, "Utility maximization for communication networks with multipath routing," *IEEE Trans. Automatic Control*, vol. 51, no. 5, pp. 766–781, May 2006.

[20] H. Khalil, *Nonlinear Systems*, Prentice Hall, 2001.

[21] F. Callier and C. Desoer, *Linear System Theory*, Springer-Verlag, 1991.